\documentclass[useAMS,referee,usenatbib]{biom}
\def\bSig\mathbf{\Sigma}

\usepackage{float}
\usepackage{amsmath}
\usepackage{graphicx,psfrag,epsf}
\usepackage{enumerate}
\usepackage{url} 
\usepackage[hidelinks]{hyperref}
\usepackage{xcolor}
\hypersetup{
    colorlinks,
    linkcolor={black!50!black},
    urlcolor={blue!80!black},
    citecolor = {black!80!black}
}
\usepackage{bm}
\usepackage{dsfont}

\usepackage[ruled,vlined]{algorithm2e}
\usepackage{multirow}
\usepackage{multicol}
\usepackage{subcaption}
\usepackage{comment}
\usepackage{adjustbox}

\DeclareMathOperator*{\argmin}{arg\,min}
\usepackage{kantlipsum,setspace}

\title[Scalable Scalar-on-Image Cortical Surface Regression]{Scalable Scalar-on-Image Cortical Surface Regression with a Relaxed-Thresholded Gaussian Process Prior}

\author{Anna Menacher$^{1,*}$\email{annamenacher@yahoo.de, anna.menacher@stats.ox.ac.uk}, 
Thomas E. Nichols$^{2,**}$\email{thomas.nichols@bdi.ox.ac.uk}, Timothy D. Johnson$^{3,***}$\email{tdjtdj@umich.edu}, and 
Jian Kang$^{3,****}$\email{jiankang@umich.edu} \\
$^{1}$Department of Statistics, University of Oxford, Oxford, UK \\
$^{2}$Nuffield Department of Population Health, University of Oxford, Oxford, UK \\
$^{3}$Department of Biostatistics, University of Michigan, Ann Arbor, Michigan, USA}

\begin{document}





\pagerange{\pageref{firstpage}--\pageref{lastpage}} 
\volume{0}
\pubyear{2024}
\artmonth{March}


\doi{}


\label{firstpage}


\begin{abstract}
In addressing the challenge of analysing the large-scale Adolescent Brain Cognition Development (ABCD) fMRI dataset, involving over 5,000 subjects and extensive neuroimaging data, we propose a scalable Bayesian scalar-on-image regression model for computational feasibility and efficiency. Our model employs a relaxed-thresholded Gaussian process (RTGP), integrating piecewise-smooth, sparse, and continuous functions capable of both hard- and soft-thresholding. This approach introduces additional flexibility in feature selection in scalar-on-image regression and leads to scalable posterior computation by adopting a variational approximation and utilising the Karhunen-Loéve expansion for Gaussian processes. This advancement substantially reduces the computational costs in vertex-wise analysis of cortical surface data in large-scale Bayesian spatial models. The model's parameter estimation and prediction accuracy and feature selection performance are validated through extensive simulation studies and an application to the ABCD study. Here, we perform regression analysis correlating intelligence scores with task-based functional MRI data, taking into account confounding factors including age, sex, and parental education level. This validation highlights our model's capability to handle large-scale neuroimaging data while maintaining computational feasibility and accuracy.
\end{abstract}

%

\begin{keywords}
Cortical surface; Gaussian process; Neuroimaging; Spatial variable selection; Variational inference.
\end{keywords}


\maketitle


%

\section{Introduction}
\label{sec: intro}


\subsection{Motivation for Analysis of Cortical Surface Data}
\label{sec: intro-cortical-surface}
Functional Magnetic Resonance Imaging (fMRI) measures brain activity and can be used to identify interactive relationships between brain areas. Measuring brain activity plays a crucial role in understanding the human brain, both in healthy individuals and in patient populations. fMRI works by measuring local changes in blood flow that reflect changes in brain activity, and is used in conjunction with cognitive tasks to identify brain regions that support those tasks. MRI modalities, such as structural or functional MRI, can be analysed in either a volumetric- or cortical surface-based representation. When it comes to the development of statistical methods for Bayesian spatial models, it can be beneficial to choose surface-based over volumetric-based representations as it provides better whole brain visualisation, dimension reduction as only the surface is of interest, removal of unnecessary tissue types, and improved representation of neurobiologcial distances \citep{mejia2020bayesian}. 

The non-invasive nature of MRI also facilitates the analysis of brain structure and function and its impact on complex human traits, such as general cognition. The relationship between human intelligence and brain features has been extensively studied focusing on cortical volume and thickness \citep{Karama2011} as well as assessing the influence of white matter integrity across the brain on the neurodevelopment of cognitive traits \citep{Muetzel2015}; for an overview, see \citet{Oxtoby2019}. Understanding the relationship between brain features and intelligence during developmental years in preadolescence can provide insights into biomarkers of resilience.

Massive datasets are currently playing a central role in neuroimaging, necessitating scalable statistical methods that can applied to thousands and, in the future, hundreds of thousands subjects \citep{Marek2022}. Many population-based health studies, such as the Adolescent Brain Cognitive Development (ABCD) study \citep{casey2018adolescent}, the UK Biobank (UKBB), and the Human Connectome Project (HCP), not only have larger sample sizes (ABCD: $\approx$~12,000 subjects, UKBB: $\approx$~100,000 subjects, HCP: $\approx$~1,200 subjects) but also include various other data sources, such as omics data, environmental factors, and mental health questionnaires. Hence, enabling the study of more complex models which respect the spatial dependency and the multi-modality of the data and are scalable to thousands of subjects. 

In this paper, we consider methods for cortical surface scalar-on-image regression, where the output is a scalar quantity and the input is an image. An example analysis is the  association of a composite intelligence score \citep{Luciana2018} with the test statistics derived from the first-level analysis of the emotional n-back task fMRI with a 2- vs. 0-back contrast \citep{casey2018adolescent}. Our main objectives are as follows: 1)~to provide scalable variable selection suitable for the analysis of a dataset of the size of the ABCD study via a variational approximation to the posterior and a Karhunen-Lo\`eve expansion of a Gaussian process, 2)~to address the non-identifiability issue inherent in scalar-on-image regressions via relaxed-thresholded Gaussian process priors, and 3)~to evaluate our method through simulation studies and apply it to an analysis of fMRI data in the ABCD study.

\subsection{Thresholded Gaussian Processes}
\label{sec: intro-thresholded-gps}
In this paper, we use boldface to denote vectors or matrices. Gaussian processes (GPs) specify the distributions of functions that map from an input $\bm{x} \in \mathds{R}^d$ to an one-dimensional output $f(\bm{x})$. They are capable of capturing flexible model structures and providing uncertainty estimates about a function, given the data. GPs can also be used for prior specifications over functions in a Bayesian model. A GP is hereby defined as a collection of random variables, such that any finite set of unique $\{\bm{x}_i\}_{i=1}^n$ implies that the output $\bm{f} = \{f(\bm{x}_1), \dots, f(\bm{x}_n)\}^T$ follows a multivariate Gaussian distribution with mean vector $\bm{m}$ and covariance matrix $\bm{C}$. 

\begin{figure}[h!]
\begin{adjustbox}{minipage=\linewidth,scale=1}

\begin{subfigure}{0.32\textwidth}
\includegraphics[width=1\linewidth]{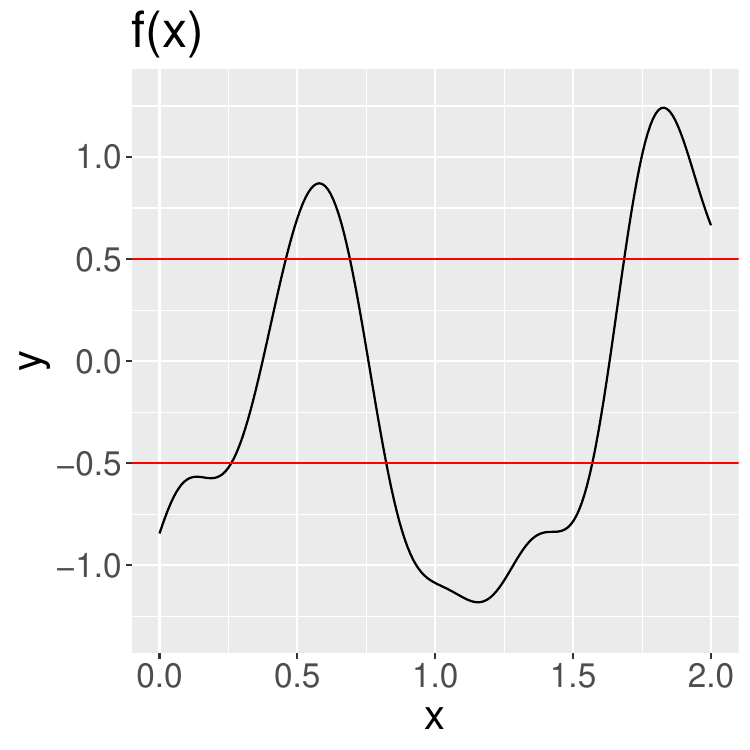}
\end{subfigure}
\begin{subfigure}{0.32\textwidth}
\includegraphics[width=1\linewidth]{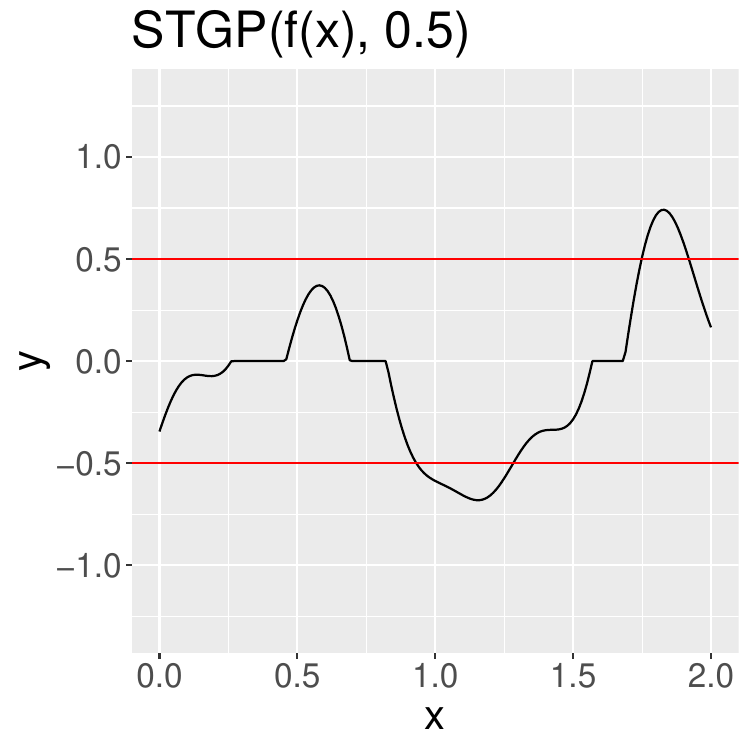}
\end{subfigure}
\begin{subfigure}{0.32\textwidth}
\includegraphics[width=1\linewidth]{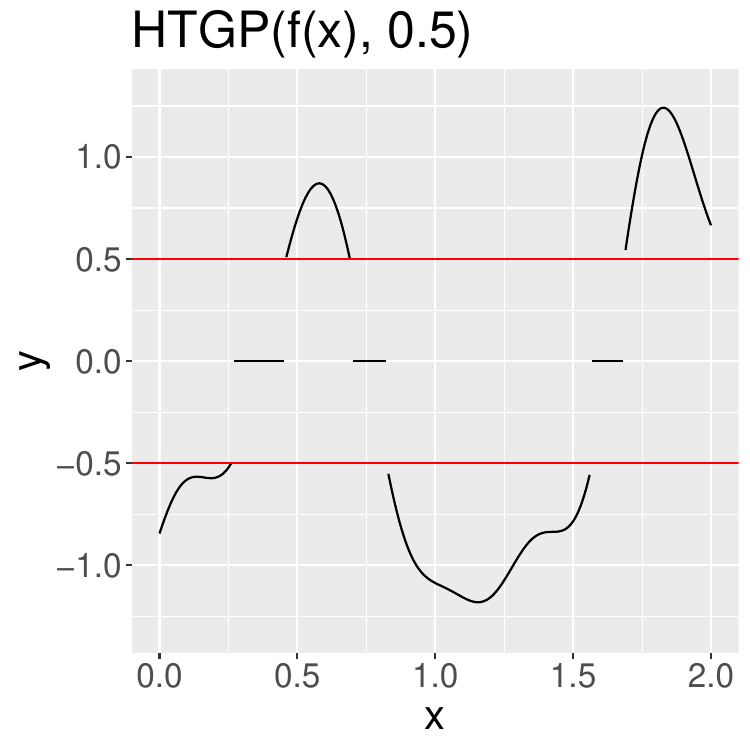}
\end{subfigure}

\begin{subfigure}{0.32\textwidth}
\includegraphics[width=1\linewidth]{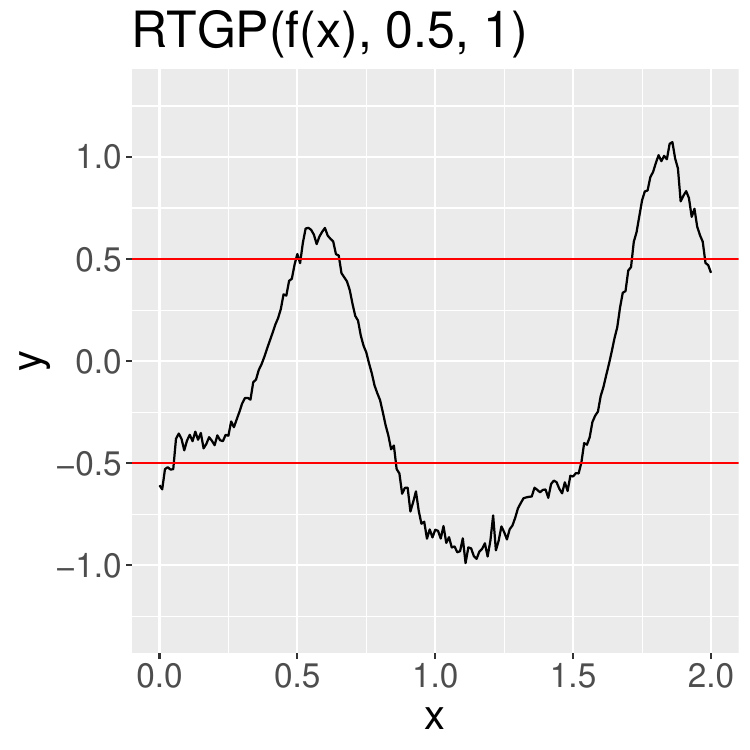}
\end{subfigure}
\begin{subfigure}{0.32\textwidth}
\includegraphics[width=1\linewidth]{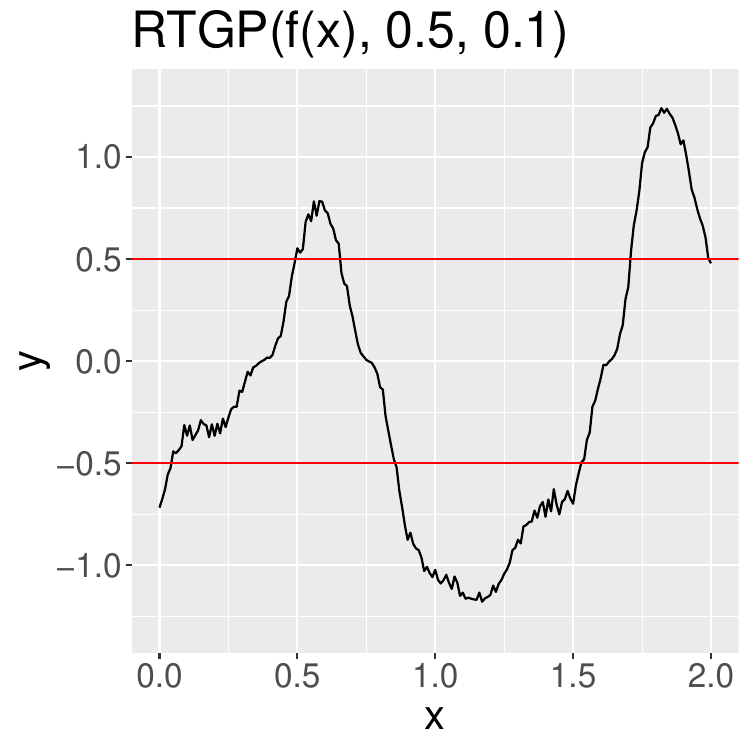}
\end{subfigure}
\begin{subfigure}{0.32\textwidth}
\includegraphics[width=1\linewidth]{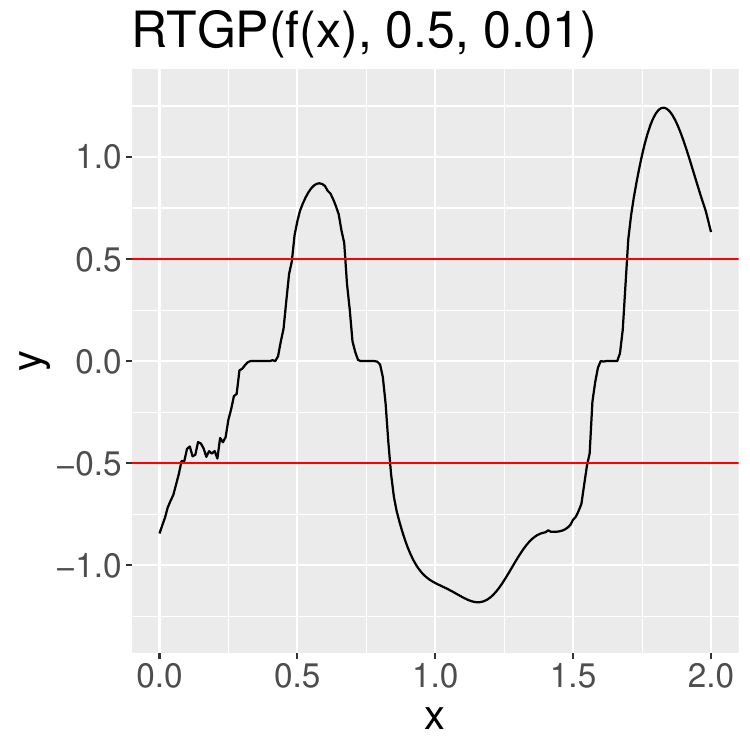}
\end{subfigure}

\end{adjustbox}
\caption[Illustration of a draw from a Gaussian Process with a squared exponential kernel $f(\bm{x})$ and passing it through various thresholding functions.]{Illustration of a draw from a Gaussian Process with a squared exponential kernel $f(\bm{x})$, passing the GP through a soft-thresholding function and a hard-thresholding function with a threshold of $\delta = 0.5$ in the top row from left to right. The bottom row depicts averaged draws from a relaxed-thresholded Gaussian process with a threshold of $\delta = 0.5$ and varying relaxing parameters $\sigma_{\alpha}^2 = \{1, 0.5, 0.1\}$. (Adapted from \citet{Li2023})}
\label{fig: gp_stgp_htgp_rtgp}
\end{figure}

A GP is fully specified by its mean function $m(\bm{x})$ and covariance function $C\{d(\bm{x}, \bm{x}')\}$, where $d(\bm{x}, \bm{x}')$ represents the distance between the two input points $\bm{x}$ and $\bm{x}'$. The choice of mean and covariance functions is crucial, as they determine the GP's properties: the mean function outlines the general trend of the function while the covariance function indicates smoothness and correlation between the input points. In  this paper, we adopt the zero function for the mean function, i.e. $m(\bm{x}) \equiv 0$ for all input points $\bm{x}$. We assume the covariance function is a stationary and positive definite kernel function that depends on the hyperparameters $\bm{\xi}$. Thus, we denote covariance function by $C_{\bm{\xi}}\{d(\bm{x}, \bm{x}')\}$. Various covariance functions are available, such as the squared exponential kernel \citep{Stein1999} or the Mat\'ern kernel \citep{Matern2013}, which can be adjusted or combined to achieve specific model characteristics. Thresholded GPs introduce sparsity into standard GPs with a smooth covariance kernel through various thresholding functions. Figure~\ref{fig: gp_stgp_htgp_rtgp} illustrates the impact of different thresholding functions, including hard-, soft-, and relaxed-thresholding, on a function sampled from a standard GP.

The hard-thresholded GP (HTGP) \citep{Shi2015} imposes sparsity on spatially varying coefficient models and hence is capable of performing variable selection. 
The HTGP is defined by placing a standard GP prior over the spatially varying coefficients~$\bm{\beta}(s)$, so that $\bm{\beta}(s) \sim \mathcal{GP}\left\{\bm{0}, \sigma_{\beta}^2 C(d(s, s'))\right\}$, where $\sigma_{\beta}^2 C\{d(s,s')\}$ is a stationary covariance function $\text{cov}(\bm{\beta}(s), \bm{\beta}(s'))$ with $\sigma_{\beta}^2$ defining the marginal variance. In a next step, the spatially varying coefficients $\beta(s_j)$ are transformed for all spatial locations $j=1, \dots, M$, where $M$ is the number of voxels or vertices within an image depending on its representation in the volume or on the surface, by applying the hard-thresholding function $g_{\text{\tiny{HTGP}}} (x, \delta) = x\ \cdot\ I(|x| > \delta)$, where the threshold $\delta > 0$ is considered a hyperparameter that controls the degree of sparsity. Overall, this prior specification adds sparsity, accounts for spatial dependence, is piecewise-smooth, and is able to detect edge effects and jumps. 

The soft-thresholded GP (STGP), introduced by~\citet{Kang2018}, integrates spatial variable selection into a Bayesian nonparamtric scalar-on-image regression model. Unlike trans-kriging \citep{Cressie1993} or Gaussian copulas \citep{Nelsen1999}, the STGP prior uniquely provides a large prior support over the class of piecewise-smooth, sparse, and continuous spatially varying regression coefficient functions. Hence, this approach enables a gradual transition between the estimation of zero and nonzero effects in neighbouring locations. Similar to the HTGP, the spatially varying coefficients $\bm{\beta}(s)$ are drawn from a standard GP $\bm{\beta}(s) \sim \mathcal{GP}\left\{\bm{0}, \sigma_{\beta}^2 C(d(s, s'))\right\}$. However, the coefficients are now passed through a soft-threshoulding function $g_{\text{\tiny{STGP}}}(x,\delta) = \text{sgn}(x)(|x| - \delta)\cdot I(|x| > \delta)$, where $\text{sgn}(x) = 1$ if $x>0$, $\text{sgn}(x) = -1$ if $x<0$, and $\text{sgn}(0) = 0$. Equivalently to the HTGP, the STGP maps coefficients near zero to exact zero and hereby induces sparsity where the thresholding parameter~$\delta > 0$ controls the level of sparsity.

\subsection{Scalable Inference for Gaussian Processes}
\label{sec: intro-scalable-inference}
\subsubsection{Spatial Process Approximations}
\label{sec: intro-spatial-process-approximations}
Spatial process models, particularly in neuroimaging applications, can be computationally intensive due to thousands of spatial locations over which a spatial process is defined. This computational complexity becomes a significant challenge in large-scale population-based health studies like the ABCD study, which encompasses data from thousands of subjects across various imaging modalities. The primary bottleneck in model fitting is the $\mathcal{O}(M^3)$ time complexity associated with matrix inversion when performing posterior inference, where $M$ is the number of spatial locations. As the number of locations increases, computation becomes increasingly challenging. Low-rank models or sparse models may provide solutions to mitigate this issue. 

The kernel convolution approximation of a spatial GP, as proposed by \citet{Higdon1999}, serves as an example of a low-rank spatial model used to ease the computational burden that spatial models usually impose. \citet{Kang2018} use this method to scale their scalar-on-image regression approach to an EEG study with a large number of spatial-temporal locations. The approach introduces a smaller subset of locations, called ``knots'', compared to the original number of locations. Hence, if the number of knots~$r$ is lower than the number of locations $M$, then computational efficiency is improved by reducing the time complexity from $\mathcal{O}(M^3)$ to $\mathcal{O}(Mr^2)$. However, the downside of low-rank models is that these approximate methods perform poorly if neighbouring locations are strongly correlated as the signal would dominate the noise \citep{Stein2014}. Moreover, when dealing with a large number of locations, achieving a satisfactory GP approximation necessitates a large number of knots, which can diminish the model's computational advantages~\citep{Datta2016}. 

Sparse methods provide an alternative solution to the approximation of large spatial processes. For example, \citet{Vecchia1988} introduce sparsity into the precision matrix in products of lower-dimensional conditional distributions. However, this Vecchia approximation for spatial processes is computationally infeasible for large-scale applications. Addressing this limitation, \citet{Katzfuss2021} develop a sparse general Vecchia approximation for Gaussian processes, enhancing its scalability to large datasets. The main issue with sparse solutions is that they primarily focus on estimation of the parameters of a covariance or precision function without providing insights about the underlying process. 

\subsubsection{Approximate Posterior Inference}
\label{sec: intro-approximate-posterior-inference}
The gold standard for parameter estimation and inference in Bayesian models with Gaussian process priors is MCMC sampling, renowned for its accurate uncertainty quantification of the spatially varying coefficients. However, the high-dimensional nature of image-based regression problems, coupled with the growing accessibility of large-scale population health studies to researchers, necessitates more scalable inference solutions. 

The need for scalability is underscored by several examples of large-scale Bayesian spatial models using thresholded GP priors. For instance, the scalar-on-image regression model with a STGP prior proposed by \citet{Kang2018} uses Metropolis-Hastings within Gibbs sampling for posterior inference.  The approach can become computationally slow if the predefined acceptance ratio of the Metropolis-Hastings algorithm is not adequately tuned or can even lead to poor parameter estimates if the mixing of the coefficients is slow and convergence is not reached within the time limit or number of iterations set by the user. Similarly, the image-based regression model with a HTGP prior proposed by \citet{Shi2015} suffers from non-conjugacy which requires the Metropolis-Hastings steps for posterior updates and uses the full conditionals for drawing samples from the remaining model parameters which exhibit conjugacy. Moreover, the scalar-on-image regression model with a RTGP prior proposed by \citet{Li2023} is able to use full conditional distributions to update the model parameters with a Gibbs sampler. While the existence of closed form conditionals increases scalability, the model is still not scalable to the number of spatial locations typically found in MRI or large-scale population health studies, such as the ABCD study.

Hence, variational inference provides a more scalable posterior inference algorithm for image-based regression problems. By transforming the task of approximating posterior densities from a sampling problem into an optimisation problem, variational inference \citep{Jordan1999, Blei2017} algorithms offer a computationally expedient alternative to MCMC methods. \citet{Roy2021} provide an extensive overview on using mean-field variational inference for high-dimensional regression scenarios with sparse priors. Another example using a variational approximation for high-dimensional regression problems has been developed by \citet{Menacher2023} for Bayesian lesion estimation with a structured spike-and-slab prior in which optimisation-based methods, such as variational inference and approximate posterior sampling, are used for posterior inference of a large-scale image-on-scalar regression problem. 

\section{Methods}
\label{sec: methods}
Throughout this work we assume that we can analyse the left and the right hemisphere of the cortex as separate analyses as each side of the cortex is responsible for its own functions or functions that occur in both hemispheres alike \citep{berlucchi1983two}. Within each analysis, we therefore make the assumption of a single hemisphere, cortical surface-based, spherical coordinate system \citep{fischl1999cortical} where we work with geodesic distances on the surface rather than Euclidean distances in the three-dimensional (3D) volume. 

Let $\mathds{S}$ denote the set of coordinates on a sphere with a known radius and $\bm{s} \subset \mathds{S}$ as a set of vertices of a single hemisphere of the cortex at which MRI data is observed. In our real data application to the ABCD study, we work with data that has been mapped from the native patient's brain space with approximately 150,000 vertices to a normalised template brain space with approximately 30,000 vertices in $\bm{s}$. For any two spatial locations $s, s' \in \mathds{S}$, we measure their distance~$d(s,s')$ with the great-circle distance between $s$ and $s'$. In theory, $\mathds{S}$ can  represent any topological surface or even a volume in a volumetric-based analysis and $d(\cdot,\cdot)$ can also represent any appropriate distance metric. The distance between $s$ and $s'$ is used in $C_{\bm{\xi}}\{d(s,s')\}$ which represents the stationary spatial correlation function defined on $\mathds{S}$ with the kernel hyperparameters $\bm{\xi}$. The correlation function $C(\cdot)$ can be any positive definite kernel function defined so that $C(0) = 1$ and $C(d) \leq 1$ for all $d>0$. 

\subsection{Model}
\label{sec: methods-model}
We propose a Bayesian scalar-on-image regression model with a relaxed-thresholded Gaussian process prior. Firstly, the scalar output $y_i$ for every subject $i=1,\dots,N$ is modelled with a Gaussian random variable 
\begin{align}
\label{eq: normal_regression}
    \left[y_i \mid \bm{x}_i, \Tilde{\bm{\beta}}(\bm{s}), \bm{\alpha}(\bm{s}), \sigma_{\epsilon}^2\right] &\sim \mathcal{N}\left\{ \beta_0 + \sum_{j=1}^M \Tilde{\beta}(s_j) I(| \alpha(s_j) | > \delta)x_{i,j}, \sigma_{\epsilon}^2 \right\}
\end{align}
with the mean expressed by the linear predictor and the variance defined as the residual noise~$\sigma_{\epsilon}^2$. Here $\bm{\alpha}(\bm{s}) = \{\alpha(s_1),\ldots, \alpha(s_M)\}^\top $, $\bm{\tilde\beta}(\bm{s}) = \{\tilde\beta(s_1),\ldots, \tilde\beta(s_M)\}^\top $ and $\bm{s}= (s_1,\ldots, s_M) $. The linear predictor contains the sum of the intercept $\beta_0$ and the linear combination of imaging covariates $x_{i,j}$ for every vertex $j=1,\ldots,M$ and thresholded spatially varying coefficients $\beta_{\text{\tiny{RTGP}}}(s_j) = \Tilde{\beta}(s_j) I(|\alpha(s_j)| > \delta)$, where $\Tilde{\beta}(s_j)$ are spatially varying coefficients, $\alpha(s_j)$ are latent variables, and $\delta$ is the threshold parameter that determines the degree of sparsity in the model. Here we only consider one imaging modality, but this approach can be extended to include multiple types of images to assist in the prediction of a scalar outcome.

Sparsity and smoothness is incorporated in the modelling of the spatially varying coefficients by placing a relaxed-thresholded Gaussian process on the coefficients which is able to capture both the HTGP and the STGP by introducing a set of latent variables $\bm{\alpha}(\bm{s})\in \mathds{R}^M$ with a ``relaxing'' parameter~$\sigma_{\alpha}^2$, the variance of the latent variables \citep{Li2023}. Thus, offering a more flexible Bayesian nonparametric prior than either the HTGP or the STGP alone, the RTGP is piecewise-smooth and sparse and models this sparsity by capturing both sparse and non-sparse patterns alike by varying the relaxing parameter. Hence, the RTGP is a more flexible and scalable prior than the STGP which cannot capture discontinuities or the HTGP which lacks scalability due to its non-continuous nature. We put a Gaussian process prior over the spatially varying coefficients 
\begin{align}
    \Tilde{\bm{\beta}}(s) &\sim \mathcal{GP}\left[\bm{0}, \sigma_{\beta}^2 C\{d(s, s')\}\right], 
\end{align}
with mean zero, stationary covariance kernel $C(\cdot)$ and parameter $\sigma_{\beta}^2$ which controls the maximum marginal variance of $\Tilde{\bm{\beta}}(\bm{s})$. For the covariance function of the GP, we choose a two parameter exponential radial basis function 
\begin{equation}
\label{eq: kernel-surface}
    C(d(s,s')) = \exp\left\{-\phi d(s,s')^\nu\right\},
\end{equation}
where $\bm{\xi} = (\phi, \nu)^T$ are the kernel hyperparameters with $\phi > 0$ and $\nu \in (0,2]$ \citep{jousse2021}. The bandwidth or inverse scale parameter $\phi$ hereby controls how rapidly the correlation decays. On the other hand, the kernel exponent $\nu$ is responsible for how much smoothness is introduced. If the kernel exponent is $\nu=2$, then the kernel is synonymous to the stationary and isotropic Gaussian kernel. Other covariance functions can also be considered; however, we choose the two parameter exponential radial basis function as Gaussian smoothing has a long history in MRI applications \citep{Whiteman2023}. 

The latent variables $\alpha(s_j)$ are drawn from a Gaussian distribution of the form
\begin{align}
    \alpha(s_j) &\sim \mathcal{N}\{\Tilde{\beta}(s_j)\}, \sigma_{\alpha}^2)\ \ \ \ \ \ \text{for\ all\ } j=1,\dots,M, 
\end{align}
where the latter is assigned for $j=1, \dots, M$ independently and $\sigma_{\alpha}^2$ defines the relaxing parameter of the latent variable $\alpha(s_j)$ that determines if the thresholding property is closer to a HTGP or a STGP. The relaxed-thresholding function does not threshold based on the values drawn from the Gaussian process $\Tilde{\bm{\beta}}(\bm{s})$ but based on the values drawn from the latent distribution for $\bm{\alpha}(\bm{s})$, so that $g_{\text{\tiny{RTGP}}}(x,\Tilde{x}, \delta) = x\cdot I(|\Tilde{x}| > \delta)$. Hence, the thresholded spatially varying coefficients are obtained by passing the coefficients $\Tilde{\beta}(s_j)$ through the relaxed-thresholding function $g_{\text{\tiny{RTGP}}}(\cdot)$ which yields $\beta_{\text{\tiny{RTGP}}}(s_j) = g_{\text{\tiny{RTGP}}}\{\Tilde{\beta}(s_j), \alpha(s_j), \delta\}$. The benefit of introducing latent variables is that the full conditional distribution of the spatially varying coefficients is now available in closed-form, as well as conjugate, which enables efficient posterior computation via a Gibbs sampler. 

The variance of the latent variables controls the independent noise added to the real process~$\Tilde{\bm{\beta}}(\bm{s})$. A small relaxing parameter $\sigma_{\alpha}^2$ preserves the mean structure of $\Tilde{\bm{\beta}}(\bm{s})$ fairly well and introduces only a slight jitter around the mean value. A large relaxing parameter $\sigma_{\alpha}^2$ on the other hand allows for more flexibility. Figure \ref{fig: gp_stgp_htgp_rtgp} highlights this behaviour by showing how the RTGP with a relaxing parameter of $\sigma_{\alpha}^2 = 0.01$ converges to the HTGP, a RTGP with a relaxing parameter of $\sigma_{\alpha}^2 = 0.1$ mimics the STGP, and lastly a RTGP with a relaxing parameter of $\sigma_{\alpha}^2 = 1$ starts to converge towards the true signal~$f(x)$. 

Furthermore, we use a Karhunen-Loève expansion which represents a standard GP with a infinite number of basis functions using Mercer's theorem
\begin{equation}
\label{eq: karhunen-loeve}
    \Tilde{\beta}(s_j) = \sum_{l=1}^{\infty} \theta_{l} \sqrt{\lambda_{l}}\psi_{l}(s_j) \ \ \ \ \ \ \ \theta_{l} \sim \mathcal{N}(0, \sigma_{\beta}^2),
\end{equation}
where $\theta_l$ are basis coefficients, $\{\lambda_l\}_{l=1}^{\infty}$ are eigenvalues in descending order and $\{\psi_l(s_j)\}_{l=1}^{\infty}$ are the corresponding orthonormal eigenfunctions. We truncate the basis expansion to approximate the Gaussian process with a finite number of parameters, so that
\begin{equation}
    \Tilde{\beta}(s_j) \approx \sum_{l=1}^L \theta_l \sqrt{\lambda_l} \psi_l(s_j),
\end{equation}
where $L$ defines the number of basis functions. The number of eigenfunctions can be determined by performing a principal components analysis where $L$ is chosen so that a certain percentage of total variation is captured, so that 
\begin{equation}
\label{eq: percentage_total_variation}
\min \left\{ l:\ \left(\sum_{k=1}^l \lambda_k\right) \Bigg / \left( \sum_{k=1}^{\infty} \lambda_k\right) \geq \kappa_L\right\},
\end{equation} 
for $\kappa_L \in (0,1)$ where a larger $\kappa_L$ provides a better approximation of the underlying GP.

For a full Bayesian hierarchical model specification, we place a Normal prior on the intercept $\beta_0 \sim \mathcal{N}(0, \sigma_{\beta_0}^2)$, a discrete Uniform prior on the threshold parameter $\delta \sim \text{Uniform}(t_{\min}, t_{\max})$ with $n_{\delta}$ options equally spaced values between $t_{\min}$ and $t_{\max}$, and Half-Cauchy priors on the standard deviation parameters, specifically $\sigma_{\beta} \sim \text{Half-Cauchy}(0,s_{\beta})$, $\sigma_{\epsilon} \sim \text{Half-Cauchy}(0,s_{\epsilon})$, and $\sigma_{\alpha} \sim \text{Half-Cauchy}(0,s_{\alpha})$. Lastly, we utilise the scale mixture representation of the Half-Cauchy priors which re-expresses the prior on the standard deviations as Inverse-Gamma priors by introducing the additional latent variables $a_{\beta}$, $a_{\epsilon}$, and $a_{\alpha}$, such that $\sigma_{\beta}^2\sim\ \text{Inverse-Gamma}(1/2,\ a_{\beta}^{-1})$, 
$\sigma_{\epsilon}^2  \sim\ \text{Inverse-Gamma}(1/2, a_{\epsilon}^{-1})$ and $\sigma_{\alpha}^2 \sim\ \text{Inverse-Gamma}(1/2, a_{\alpha}^{-1})$
along with $a_{\beta} \sim\ \text{Inverse-Gamma}(1/2,\ s_{\beta}^{-2})$,
$a_{\epsilon} \sim\ \text{Inverse-Gamma}(1/2,\ s_{\epsilon}^{-2})$ and 
$a_{\alpha} \sim\ \text{Inverse-Gamma}(1/2,s_{\alpha}^{-2})$,
where $s_{\beta}^2$, $s_{\epsilon}^2$, and $s_{\alpha}^2$ are considered hyperparameters of the model. We use scale-rate parameterisation of the Inverse-Gamma distribution.

\subsection{Posterior Computation}
\label{sec: methods-posterior-computation}
Previous scalar-on-image regression models with HTGP or STGP prior distributions on the spatially varying coefficients require MCMC sampling for posterior computation, a method that does not scale well to large-scale studies and high-dimensional voxel- or vertex-wise analyses commonly performed in neuroimaging applications. Specifically, both thresholded coefficient models rely on Metropolis-Hastings within Gibbs sampling with a kernel convolution approximation for spatial GPs for estimating model parameters. While this approach is computationally more efficient than other MCMC alternatives, it is nonetheless an approximation and moreover the quality of parameter estimation is sensitive to the choice of proposal distributions. Our contributions to increase the scalability for posterior computation are two-fold: 1) by deriving a Gibbs sampler to perform more efficient MCMC sampling by utilising the conjugacy within the RTGP model (derivations are provided in Section 1 and simulation study results in Section 3 and 5.5 of Web Appendix A), and 2) by using variational optimisation algorithms rather than MCMC sampling to acquire posterior estimates. 

The derivation of a Gibbs sampler for the set of model parameters, given by 
\begin{equation}
\label{eq: full_set_of_model_params}
    \bm{\Omega} = \left\{\beta_0, \bm{\theta}, \bm{\alpha}(\bm{s}),\allowbreak \delta, \sigma_{\epsilon}^2, \sigma_{\beta}^2, \sigma_{\alpha}^2, a_{\epsilon}, a_{\beta}, a_{\alpha} \right\},
    \end{equation}
of our Bayesian scalar-on-image regression model with a RTGP prior provides the foundation for more scalable posterior inference solutions. Specifically, we propose the usage of mean-field variational inference (MFVI) \citep{Blei2017} rather than MCMC sampling to reduce the computational cost of this modelling framework. Generally, variational inference requires the full joint distribution of the Bayesian spatial regression model which consists of the likelihood $p(\bm{y} | \bm{X}, \Tilde{\bm{\beta}}(\bm{s}), \bm{\alpha}(\bm{s}), \delta, \beta_0, \sigma_{\epsilon}^2)$ and the joint prior distribution $p(\beta_0, \bm{\theta}, \bm{\alpha}(\bm{s}), \delta, \sigma_{\epsilon}^2,$ $\sigma_{\beta}^2, \sigma_{\alpha}^2, a_{\epsilon}, a_{\beta}, a_{\alpha})$ which is expressed through its prior conditional distributions. 
In our model, the factorisation of the variational distributions for MFVI is expressed by 
\begin{align*}
    &q(\beta_0, \bm{\theta}, \bm{\alpha}(\bm{s}), \delta, \sigma_{\epsilon}^2, \sigma_{\beta}^2, \sigma_{\alpha}^2, a_{\epsilon}, a_{\beta}, a_{\alpha}) \\
    &= q(\beta_0) \times q(\bm{\theta}) \times \prod_{j=1}^M q\{\alpha(s_j)\}\times q(\delta) \times 
     q(\sigma_{\epsilon}^2) \times q(\sigma_{\beta}^2)\times q(\sigma_{\alpha}^2) \times q(a_{\epsilon}) \times q(a_{\beta}) \times q(a_{\alpha}).
\end{align*}
Optimisation in variational inference then aims at minimising the Kullback-Leibler (KL)-divergence between the posterior distribution and the variational candidate distribution and is defined by
\begin{equation}
    q^*(\bm{\omega}) = \argmin_{q(\bm{\omega}) \in \mathcal{Q}} \text{KL}\ \{q(\bm{\omega})\  ||\  p(\bm{\omega} | \bm{y})\},
\end{equation}
where $q^*(\bm{\omega})$ for the model parameter $\bm{\omega} \in \bm{\Omega}$ provides the best approximation to the difference between the posterior and candidate distribution. However, the minimisation of the KL-divergence requires the estimation of the log-marginal of the data which is often infeasible to compute. To avoid this issue, the evidence lower bound (ELBO) is the quantity which is optimised over in lieu of the KL-divergence as it does not require the computation of the model evidence. The ELBO is defined by 
\begin{equation}
    \mathcal{L}(q) 
    \geq \mathds{E}_{q(\bm{\Omega})}\left[\ln\left\{p(\bm{Y},\bm{X},\bm{\Omega})\right\}\right] - \mathds{E}_{q(\bm{\Omega})}\left[\ln\left\{q(\bm{\Omega})\right\}\right].
\end{equation}

The variational density $q(\bm{\omega}_k)$, note we index the model parameters in the set $\bm{\Omega}$ with $k$, is derived by the exponentiated expected log of the complete conditional given the other model parameters $\bm{\Omega}_{-k}$ and the data which is defined as $q_k(\bm{\omega}_k) \propto \exp\left(\mathds{E}_{-k}[\ln\{p(\bm{\omega}_k |\bm{\Omega}_{-k}, \bm{X})\}]\right)$ where the expectation is over the fixed variational density of other variables, given by $\prod_{\ell \neq k} q_{\ell}(\bm{\omega}_{\ell})$. To provide a concrete example, we show how the full conditional distributions, derived for the Gibbs sampler, enable the derivation of the variational distributions for the basis coefficients $\bm{\theta}$ of the Karhunen-Loève expansion and the latent variables $\bm{\alpha}(\bm{s})$ which jointly provide the thresholded spatially varying coefficient values $\beta(s_j)$ for all $j=1,\dots, M$. 

Due to the Karhunen-Loève expansion, we no longer estimate the spatially varying coefficients themselves but rather acquire the variational posterior of the basis coefficients
\begin{align}
    \ln\{q^*(\bm{\theta})\} &\propto \mathds{E}_{q(\beta_0, \bm{\alpha}(\bm{s}), \delta, \sigma_{\epsilon}^2, \sigma_{\beta}^2, \sigma_{\alpha}^2)}[\ln\{p(\bm{y} | \beta_0, \bm{\theta}, \bm{\alpha}(\bm{s}), \delta, \sigma_{\epsilon}^2)p(\bm{\theta} | \sigma_{\beta}^2) p(\bm{\alpha}(\bm{s}) | \bm{\theta}, \sigma_{\alpha}^2)\}], 
\end{align}
where the variational posterior $q^*(\bm{\theta}) = \mathcal{N}(\bm{\theta};\ \bm{\mu}_{\theta},\ \bm{\Sigma}_{\theta})$ is a Normal distribution with mean~$\bm{\mu}_{\theta}$ and covariance~$\bm{\Sigma}_{\theta}$. The regularisation of the basis coefficients comes in through prior on the basis coefficients $\bm{\theta}$ as well as the prior on the latent variables $\bm{\alpha}(\bm{s})$. The posterior estimates of~$\bm{\theta}$ then provide information on the spatially varying coefficients $\bm{\Tilde{\beta}(s)}$ by plugging the posterior quantity of interest in Equation (\ref{eq: karhunen-loeve}), so that the spatially varying coefficients $\Tilde{\beta}(s_j) \approx \sum_{l=1}^L \theta_l \sqrt{\lambda_l} \Psi_l(s_j)$ are obtained.

The variational posterior of the latent variables $\alpha(s_j)$ for each $j=1, \dots, M$ is acquired by a mixture of truncated normal distributions
\begin{align}
    \ln[q^*\{\bm{\alpha}(\bm{s})\}] &\propto \mathds{E}_{q(\beta_0, \delta, \theta, \sigma^2_{\epsilon},\sigma^2_{\alpha})} [ \ln\{p(\bm{y} \mid \beta_0, \bm{\alpha}(\bm{s}), \bm{\theta}, \delta, \sigma^2_{\epsilon}) p(\bm{\alpha}(\bm{s}) \mid \bm{\theta}, \sigma^2_{\alpha})\}] \\
     q^*\{\alpha(s_j)\} &= w^*_{-1,j}\ \text{Truncated-Normal}_{(-\infty, -\delta)}\left(\mathds{E}_{q(\theta)}[\Tilde{\beta}(s_j)], 1/\mathds{E}_{q(\sigma^2_{\alpha})}[\sigma_{\alpha}^{-2}] \right) + \\
    &\ \ \ \ w^*_{0,j}\ \text{Truncated-Normal}_{(-\delta, \delta)}\left(\mathds{E}_{q(\theta)}[\Tilde{\beta}(s_j)], 1/\mathds{E}_{q(\sigma^2_{\alpha})}[\sigma_{\alpha}^{-2}] \right) + \nonumber \\
&\ \ \ \ w^*_{1,j}\ \text{Truncated-Normal}_{(\delta, \infty)}\left(\mathds{E}_{q(\theta)}[\Tilde{\beta}(s_j)], 1/\mathds{E}_{q(\sigma^2_{\alpha})}[\sigma_{\alpha}^{-2}] \right), \nonumber 
\end{align}
where the mean is given by the spatially varying coefficients $\Tilde{\beta}(s_j) = \sum_{l=1}^L \theta_l \sqrt{\lambda_l} \Psi_l(s_j)$ and the variance is given by the prior variance of the latent variables $\alpha(s_j)$. The mixture weights $\{w^*_{-1,j}, w^*_{0,j}, w^*_{1,j}\}$ are required to sum to 1. For further detail on the weights, we refer the reader to Web Appendix~A, Section 2.4. 

Lastly, the mean-field coordinate ascent variational inference (CAVI) algorithm \citep{Bishop2006} iterates through the steps of determining a suitable variational distribution $q(\bm{\omega})$ and of updating its variational parameters, while keeping the other parameters in~$\bm{\Omega}$ fixed, until a convergence criterion is satisfied. All derivation details and information on initialisation of the variational inference algorithm can be found in Web Appendix~A, Section~2. 

\section{Simulation Study}
\label{sec: results} 
We carry out simulation studies to evaluate the performance of the proposed method, i.e. RTGP estimated via variational inference. We compare our model to simple baseline models, including a Bayesian regression model with a Normal prior (BR + Normal) and a Horseshoe prior (BR + Horseshoe) on the image coefficients~$\beta(s_j)$ for each $j=1,\dots,M$, and more complex baseline models, including a Gaussian process regression by transforming the input image data with the bases, consisting of the eigenfunctions and eigenvalues of the kernel decomposition, and placing a Normal prior (GPR + Normal) and a Horseshoe prior (GPR + Horseshoe) on the basis coefficients $\theta_l$ for each $l=1,\dots,L$. We also compare our model against baseline frequentist approaches, such as Ridge and LASSO regression. We provide a more detailed overview of the model setup of the baseline models in Web Appendix~A, Section~5.2. For the simulation study we use a number of basis functions $L=100$. We also refer the reader to Web Appendix A, Section 5.3 for further implementation details.

For a comprehensive evaluation of different methods, we evaluate  parameter estimate accuracy via absolute bias $\beta_{\mathrm{bias}} = M^{-1}\sum_{j=1}^M |\beta(s_j) - \hat{\beta}(s_j)|$ and mean squared error (MSE) $\beta_{\mathrm{MSE}} = M^{-1}\sum_{j=1}^M \{\beta(s_j) - \hat{\beta}(s_j)\}^2$, and assess the prediction accuracy via predictive $R^2 = \text{cor}(y, \hat{y})^2$ and MSE $y_{\mathrm{MSE}} = N^{-1}\sum_{j=1}^N \{y - \hat{y}\}^2$, and the feature selection accuracy by assessing true positive (TP), false positive (FP), true negative (TN), and false negative (FN) discoveries in the following measures: (1) sensitivity/true positive rate (TPR = $\text{TP}/(\text{TP} + \text{FN})$), (2) true discovery rate (TDR = $\text{TP}/(\text{TP} + \text{FP})$), (3) specificity/1 - false positive rate (FPR = $\text{FP}/(\text{FP} + \text{TN})$), and (4) false discovery rate (FDR = $\text{FP}/(\text{FP} + \text{TP})$). The feature selection uncertainty quantification for RTGP can be obtained by thresholding the expected value $\mathds{E}[I(|\alpha(s_j)| > \delta)]$ for each vertex $j=1,\dots,M$ according to the median probability model \citep{Barbieri2004}
 and for the other Bayesian approaches (BR + Normal, BR + Horseshoe, GPR + Normal, GPR + Horseshoe) by determining whether or nor 0 is included in the 95\%-HPDI (highest posterior density interval). 
 
We consider different simulation settings by varying the sample size $N \in \{500; 1,000; 2,000\}$. The test dataset in the respective simulation studies always consists of 1,000 subjects. In order to ensure the robustness of our results, we evaluate all results on 10 replicated datasets and average their results. While our focus is on the analysis of cortical surface data, we also perform simulation studies for the analysis of volumetric data where we use a modified squared exponential kernel with Euclidean distance, see Web Appendix A, Section 3. The data generating process of this surface-based simulation study is deliberately different from the model in order to provide a fair comparison between RTGP and other approaches, see Web Appendix A, Section 5.1 for a detailed overview. The true effect map $\bm{\beta}(\bm{s})$ is derived and subsampled to $M \approx 2000$ vertices from a real image-on-scalar regression problem, the input images $\bm{x}_i$ are drawn from a GP, and the scalar outputs are generated by $y_i = \beta_0 + \bm{x}_i^T \bm{\beta}(\bm{s}) + \epsilon_i$, where the true intercept is $\beta_0 = 2$ and the random error is drawn from $\epsilon_i \sim \mathcal{N}(0, \sigma^2 = 0.2)$. We focus on the results from the sample size scenario $N=500$ within the main paper and refer the reader to Web Appendix, Section 5.4 for further results.

\begin{table}[h!]
    \centering
    \begin{tabular}{ lrrrr } 
  \hline
   \textbf{Parameter Estimates} & \textbf{Bias} & \textbf{MSE} \\
  \hline
RTGP & \textbf{6.56} (1.97) & \textbf{3.23} (0.69) \\ 
  GPR + Normal & 11.77 (0.40) & 3.37 (0.14) \\ 
  GPR + Horseshoe & 12.48 (1.67) & 3.74 (0.58) \\ 
  BR + Normal & 10.81 (0.19) & 3.28 (0.11) \\ 
  BR + Horseshoe & 8.11 (0.31) & 19.14 (4.16) \\ 
  Ridge & 11.50 (0.33) & 3.47 (0.16) \\ 
  LASSO & 7.89 (0.43) & 22.24 (4.74) \\  
   \hline
  \textbf{Prediction}    & \textbf{R}$^2$ \textbf{(train)} & \textbf{MSE (train)} & \textbf{R}$^2$ \textbf{(test)} & \textbf{MSE (test)}\\
  \hline
RTGP & 49.89 (1.41) & 3.92 (0.31) & 39.38 (2.33) & 4.86 (0.30) \\ 
  GPR + Normal & 55.65 (2.23) & 3.37 (0.30) & 35.32 (2.21) & 4.82 (0.23) \\ 
  GPR + Horseshoe & 53.66 (3.00) & 3.58 (0.36) & 34.21 (2.37) & 4.88 (0.22) \\ 
  BR + Normal & \textbf{57.18} (2.65) & \textbf{3.32} (0.33) & 35.60 (2.41) & 4.78 (0.21) \\ 
  BR + Horseshoe & 53.03 (3.04) & 3.54 (0.34) & \textbf{41.07} (2.56) & \textbf{4.37} (0.19) \\ 
  Ridge & 56.95 (2.69) & 3.35 (0.32) & 35.48 (2.35) & 4.79 (0.22) \\ 
  LASSO & 52.50 (3.35) & 3.67 (0.41) & 40.22 (2.51) & 4.45 (0.15) \\
  \hline
     \textbf{Variable selection}     & \textbf{TPR} & \textbf{TDR} & \textbf{FPR} & \textbf{FDR}\\
  \hline
RTGP & 75.22 (7.03) & \textbf{56.85} (17.96) & \textbf{6.23} (5.78) & \textbf{44.15} (17.96) \\ 
  GPR + Normal & \textbf{89.00} (2.98) & 34.57 (8.68) & 9.80 (4.22) & 65.43 (8.68) \\ 
  GPR + Horseshoe & 87.00 (3.40) & 26.98 (7.69) & 13.93 (6.35) & 73.02 (7.69) \\ 
\hline
\vspace{0.5cm}
    \end{tabular}
    \caption{Evaluation of parameter estimate, prediction, and variable selection results (Monte Carlo standard error) for a simulation study setting with sample size $N=500$, see Web Appendix A, Section 5.4 for further sample size settings, for our model RTGP and the baseline models GPR + Normal, GPR + Horseshoe, BR + Normal, BR + Horseshoe, Ridge and LASSO. The values are multiplied by a scaling factor of $10^2$ for clarity.}
    \label{tab: simstudy-n500-results}
\end{table}

\begin{figure}[H]
\begin{adjustbox}{minipage=\linewidth,scale=1}
\rotatebox[origin=l]{90}{\ \  \ \ \ \ \  \ \ \   (1) Parameter Map}
\begin{subfigure}{0.23\textwidth}
\caption{True beta}
\includegraphics[width=1\linewidth]{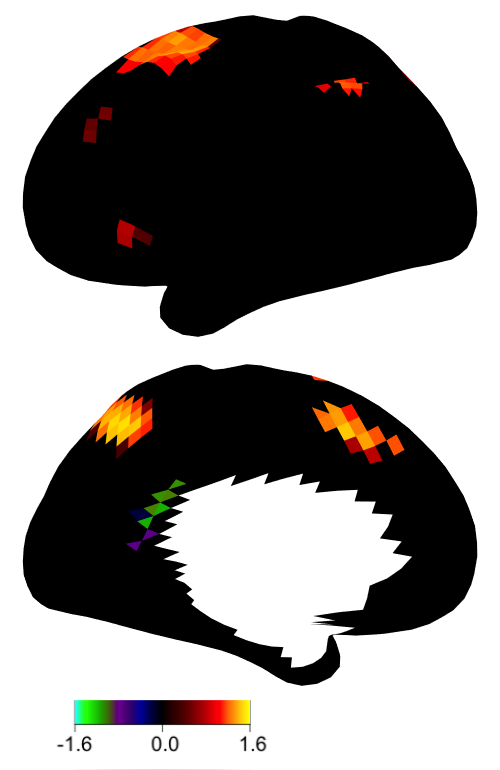}
\end{subfigure}
\begin{subfigure}{0.23\textwidth}
\caption{RTGP}
\includegraphics[width=1\linewidth]{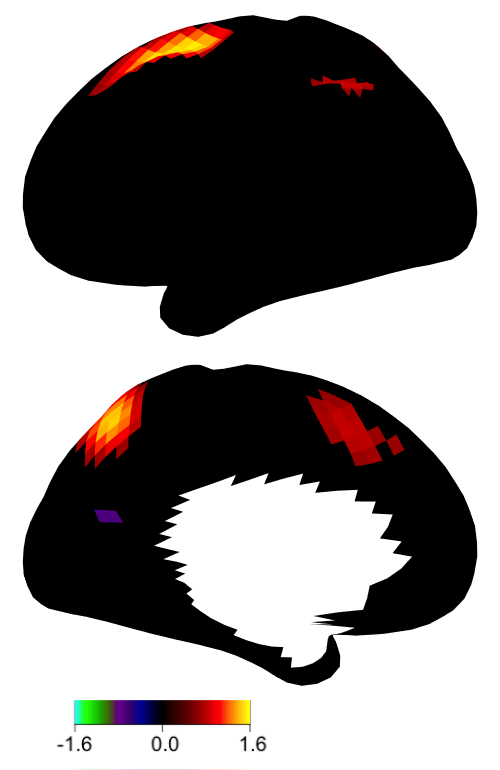}
\end{subfigure}
\begin{subfigure}{0.23\textwidth}
\caption{GPR + Normal}
\includegraphics[width=1\linewidth]{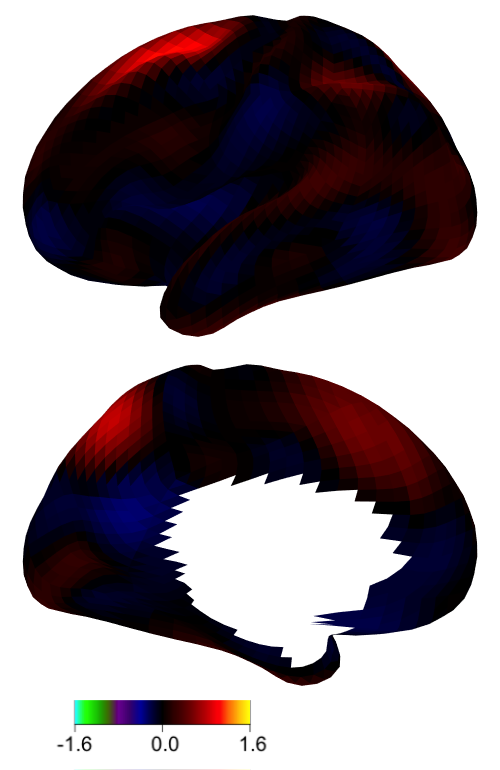}
\end{subfigure}
\begin{subfigure}{0.23\textwidth}
\caption{GPR + Horseshoe}
\includegraphics[width=1\linewidth]{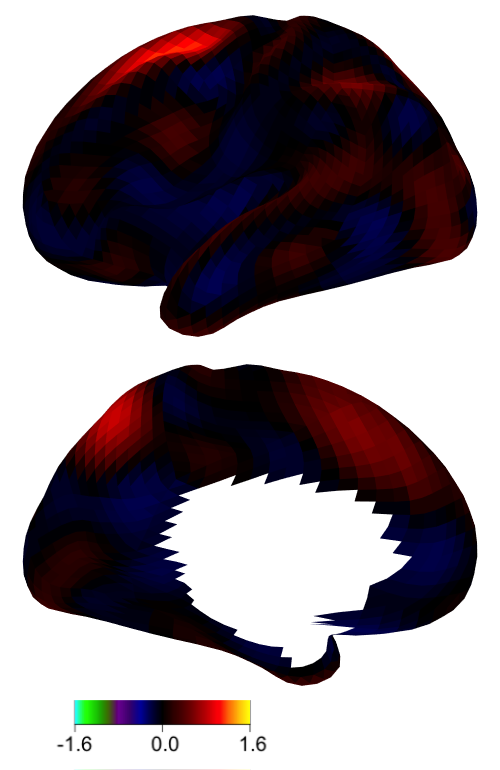}
\end{subfigure}

\rotatebox[origin=l]{90}{\ \  \ \ \ \ \  \ \  (1) Parameter Map}
\begin{subfigure}{0.23\textwidth}
\caption{BR + Normal}
\includegraphics[width=1\linewidth]{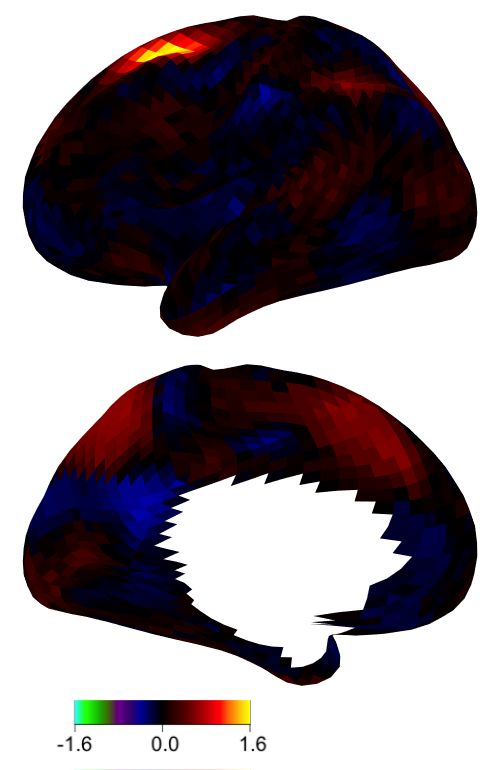}
\end{subfigure}
\begin{subfigure}{0.23\textwidth}
\caption{BR + Horseshoe}
\includegraphics[width=1\linewidth]{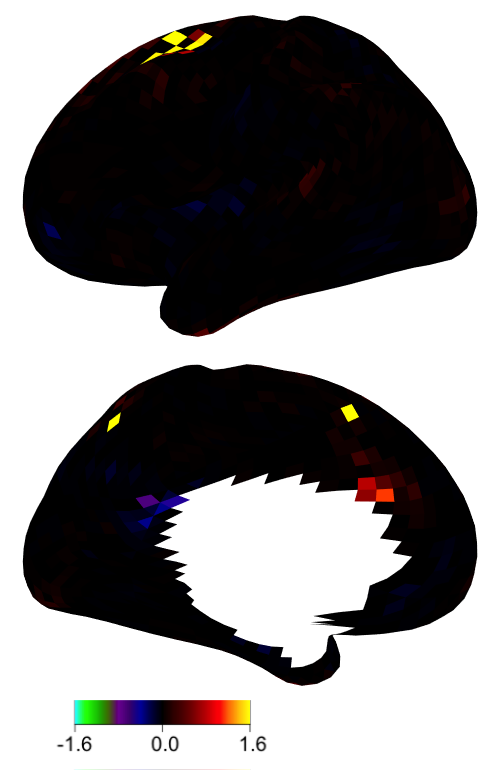}
\end{subfigure}
\begin{subfigure}{0.23\textwidth}
\caption{Ridge}
\includegraphics[width=1\linewidth]{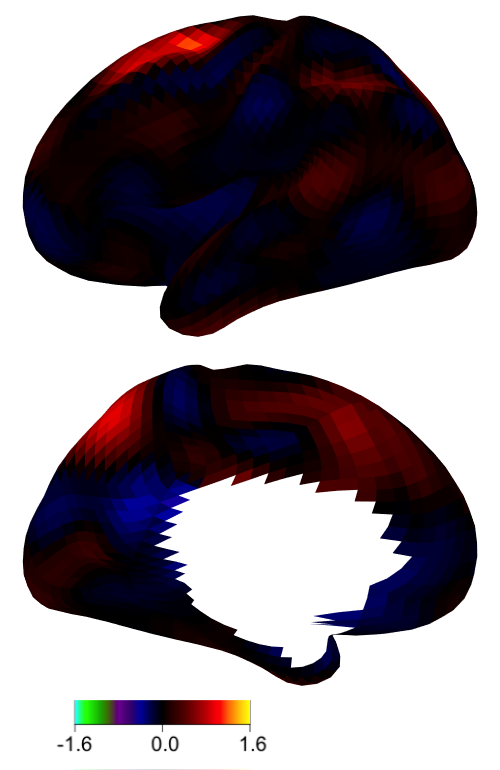}
\end{subfigure}
\begin{subfigure}{0.23\textwidth}
\caption{LASSO}
\includegraphics[width=1\linewidth]{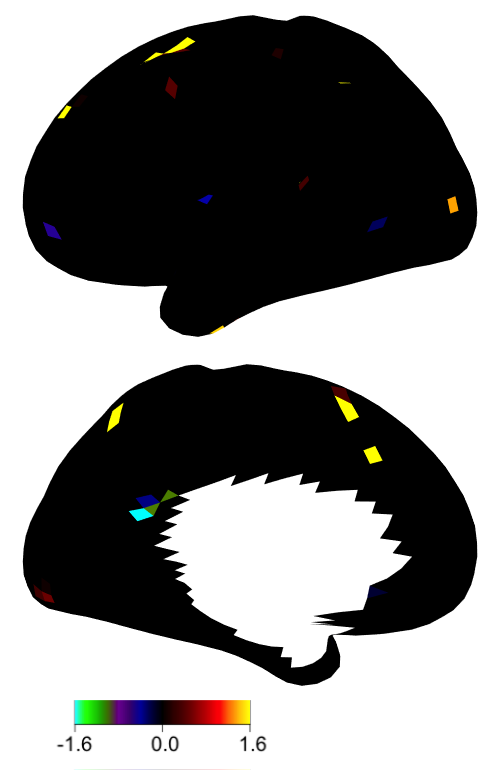}
\end{subfigure}

\rotatebox[origin=l]{90}{\ \ \  \ \ \ \ \ (2) Significance Map}
\begin{subfigure}{0.23\textwidth}
\caption{True significance}
\includegraphics[width=1\linewidth]{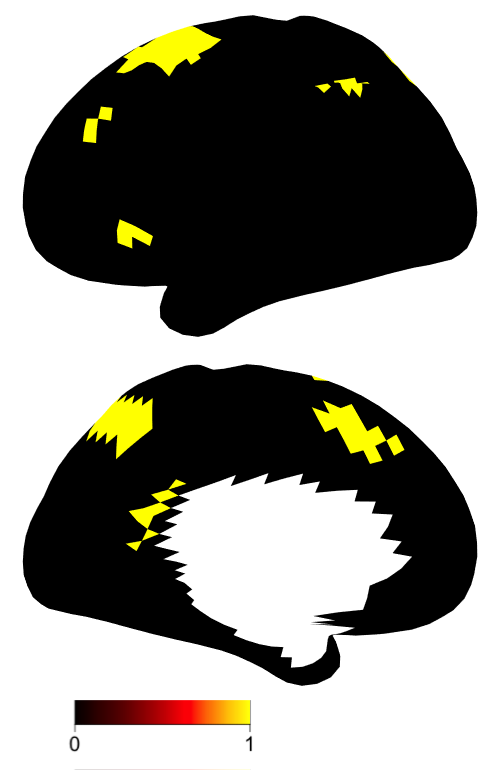}
\end{subfigure}
\begin{subfigure}{0.23\textwidth}
\caption{RTGP}
\includegraphics[width=1\linewidth]{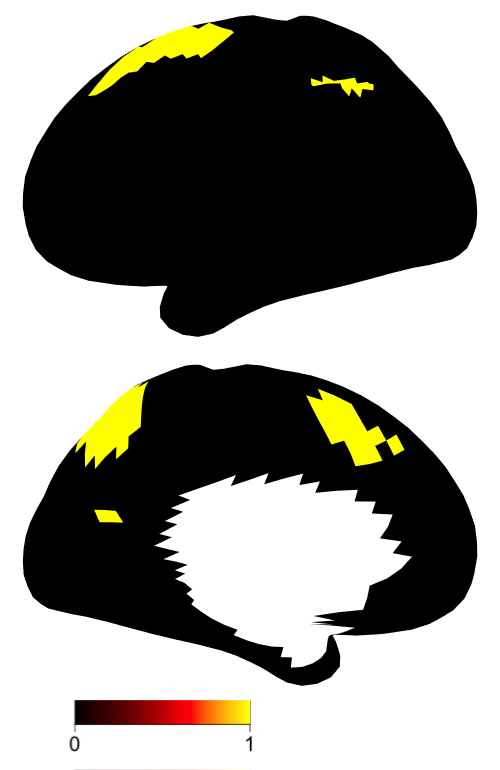}
\end{subfigure}
\begin{subfigure}{0.23\textwidth}
\caption{GPR + Normal}
\includegraphics[width=1\linewidth]{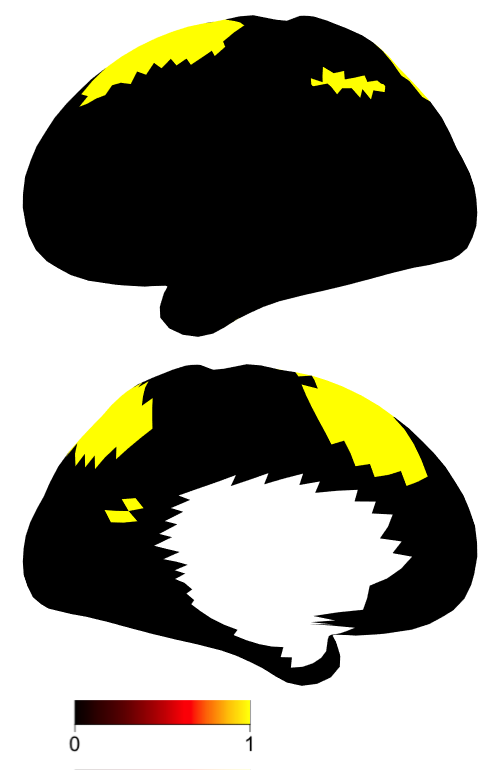}
\end{subfigure}
\begin{subfigure}{0.23\textwidth}
\caption{GPR + Horseshoe}
\includegraphics[width=1\linewidth]{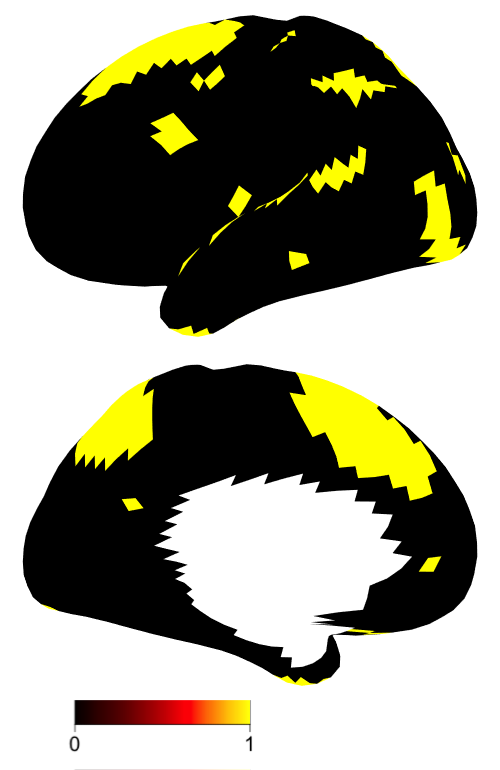}
\end{subfigure}
\end{adjustbox}
\caption{Simulation study setting (N = 500): (1) Comparison of the (a) true parameter map with the estimated parameter maps from (b) our RTGP model and (c)-(h) the other baseline approaches. Note that the we limit the colourbar to the range of the true parameter values which lie between -1.6 to 1.6 for a fair comparison of models. However, the (f) BR + Horseshoe model and (h) the frequentist LASSO regression both exhibit far more extreme values which lie outside the plotted range. (2) Comparison of the (i) true significance map with the estimated binary significance maps for (j) our RTGP model and the (k)-(l) GP regressions. We omit the significance maps of the other models because they either show now significance across the cortex (BR + Normal, BR + Horseshoe), significance cannot be obtained (Ridge) or significance is too speckled (LASSO).}
\label{fig: simstudy-parameter-and-inference}
\end{figure}

Firstly, we evaluate parameter estimate accuracy and find that our model RTGP has consistently lower absolute bias and MSE for all estimated sample sizes than any of the other baseline approaches as shown in Table 5 in the Web Appendix A. Figure~\ref{fig: simstudy-parameter-and-inference} (1) supports that RTGP outperforms the other baseline approaches with respect to parameter estimation where RTGP is able to not only capture the true effect size magnitude much better than the other models but also is able to pick up where the spatial signal is located far more accurately than the other approaches which suffer from oversmoothing in the case of GPR + Normal, GPR + Horseshoe, BR + Normal, and Ridge regression. On the other hand, BR + Horseshoe and LASSO regression exhibit very localised, speckled and extreme estimated parameter values as these models ignore the spatial dependency between neighbouring vertices and put extreme weight on a single vertex rather than smoothing the effect. In doing so both models favour solutions which are sparse in nature and aim at preventing overfitting; however, in doing so accurate parameter estimation and variable selection is jeopardised while out-of-sample prediction still yields comparable or superior results (see Table~\ref{tab: simstudy-n500-results}).

Secondly, we find that the out-of-sample prediction accuracy are comparable across all models which is common in scalar-on-image regression problems which suffer from non-identifiability if no model constraints are imposed. The lowest possible MSE for example is given by the residual noise which is known in the simulation study with $\sigma_{\epsilon}^2 = 0.04$ (multiplied with scaling factor the true residual noise variance is $\sigma_{\epsilon}^2 = 4$). In the simulation studies, we also observe that approaches which have poor variable selection results, such as BR + Horseshoe for example with no observed activation, exhibit better predictive performance with a R$^2$ (test) = 41.07\% and MSE (test) = 4.37 for a sample size of $N=500$ whereas our RTGP model has a R$^2$ (test) = 39.38\% and MSE (test) = 4.86 (see Table~\ref{tab: simstudy-n500-results}). Hence, showcasing that many combinations of spatial locations can be predictive of our outcome of interest; however, depending on what a researcher is interested in it can be favourable to choose a model which has slightly worse out-of-sample predictive results with respect to R$^2$ and MSE but has superior variable selection results with lower false positive activation compared to the other approaches.

Thirdly, the evaluation of feature selection can only be performed for a comparison of RTGP to the baseline models that involve a Gaussian process regression (GPR + Normal, GPR + Horseshoe). For the other approaches, significance can either not be obtained as in the case of the frequentist approach (Ridge), significance is too speckled (LASSO) or no significant result is observed for the Bayesian regression models which place a Normal or Horseshoe prior over the image coefficients (BR + Normal, BR + Horseshoe). Table~\ref{tab: simstudy-n500-results} in the main paper and Table~7 in the Web Appendix A show that GPR + Normal / Horseshoe has higher sensitivity across all sample sizes than RTGP, e.g. with TPR = 87.00 for GPR + Horseshoe for N=500, but suffers from identifying many false positive clusters, see Figure~\ref{fig: simstudy-parameter-and-inference} (2), leading to low specificity, e.g. with FPR = 13.93. On the other hand, our RTGP has only comparable sensitivity results but otherwise outperforms the baseline Gaussian process regressions with respect to TDR, FPR, and FDR for all sample sizes. For example, in the case of sample size $N=500$ the sensitivity of RTGP is TPR = 75.22 but the specificity is FPR = 6.23 where the results are even more pronounced for increasing sample sizes. Figure~\ref{fig: simstudy-parameter-and-inference} (2) support the results described above where we can observe accurate spatially localised activation compared to the truth for RTGP whereas both GPR + Normal and GPR + Horseshoe are able to localise the major clusters of significance but significance is determined far beyond the borders of the true cluster of significance leading to false positive activation. Additionally, GPR + Horseshoe suffers from false positive activation across the cortex and seems to identify far more clusters as significant than are actually present in the true significance map.

\section{ABCD Study Application}
\label{sec: results-abcd}
In the real data application on the ABCD study, we now highlight how our model scales to far more complex problems with larger sample sizes and number of image locations than previously exposed to in the simulation setting. The ABCD study consists of a diverse sample size of approximately 12,000 participants aged between 9-10 years old. We utilise a preprocessed population subset of the ABCD study of 4,170 subjects which has the emotional n-back task fMRI data (specifically the 2- vs. 0-back contrast) and cognition score available without any missing values alongside several confounding variables, such as sex, age, race/ethnicity, parents' educational level, family income, parents' marital status, and imaging site ID, see Web Appendix A, Section 6.1 for an overview, inclusion criteria, and preprocessing protocol. We also split our dataset in training, validation and test data with a ratio of $8 / 1 / 1$ and absolute sample sizes of $N_{train} = 3,136$, $N_{val} = 517$, and $N_{test} = 517$. In this application, we showcase how our model RTGP behaves in comparison to the baseline approaches introduced in the previous section. We also use the same evaluation criteria as specified in Section~\ref{sec: results}. Contrary to the simulation study, we use $L=800$ which captures approximately 30\% of the total variation determined through Equation~(\ref{eq: percentage_total_variation}), see Web Appendix A, Section~6.3. 

\begin{figure}[H]
\begin{adjustbox}{minipage=\linewidth,scale=0.85}
\begin{center}

\begin{subfigure}{0.45\textwidth}
\caption{RTGP (unthresholded)}
\includegraphics[width=1\linewidth]{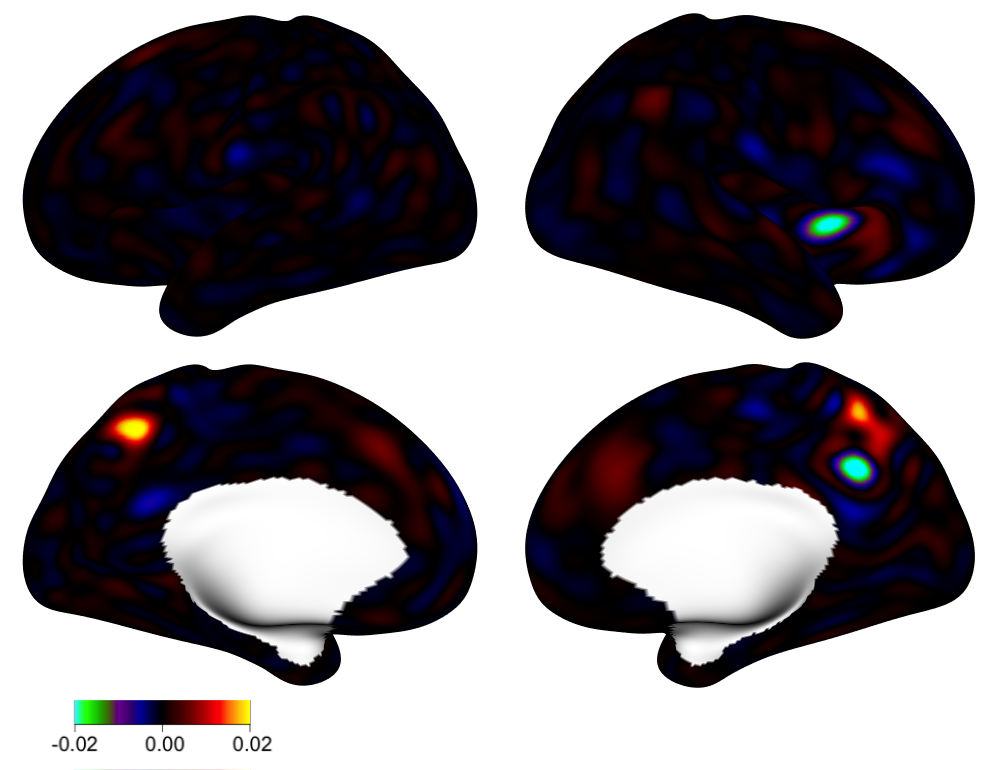}
\end{subfigure}
\begin{subfigure}{0.45\textwidth}
\caption{RTGP (thresholded)}
\includegraphics[width=1\linewidth]{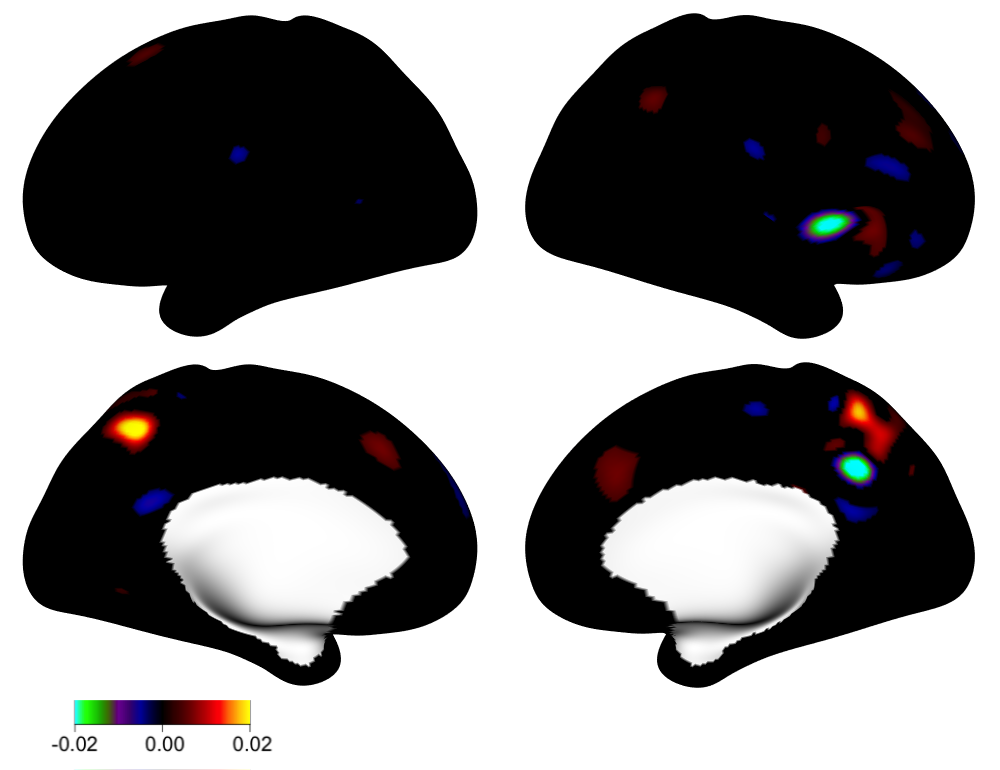}
\end{subfigure}

\begin{subfigure}{0.45\textwidth}
\caption{GPR + Normal}
\includegraphics[width=1\linewidth]{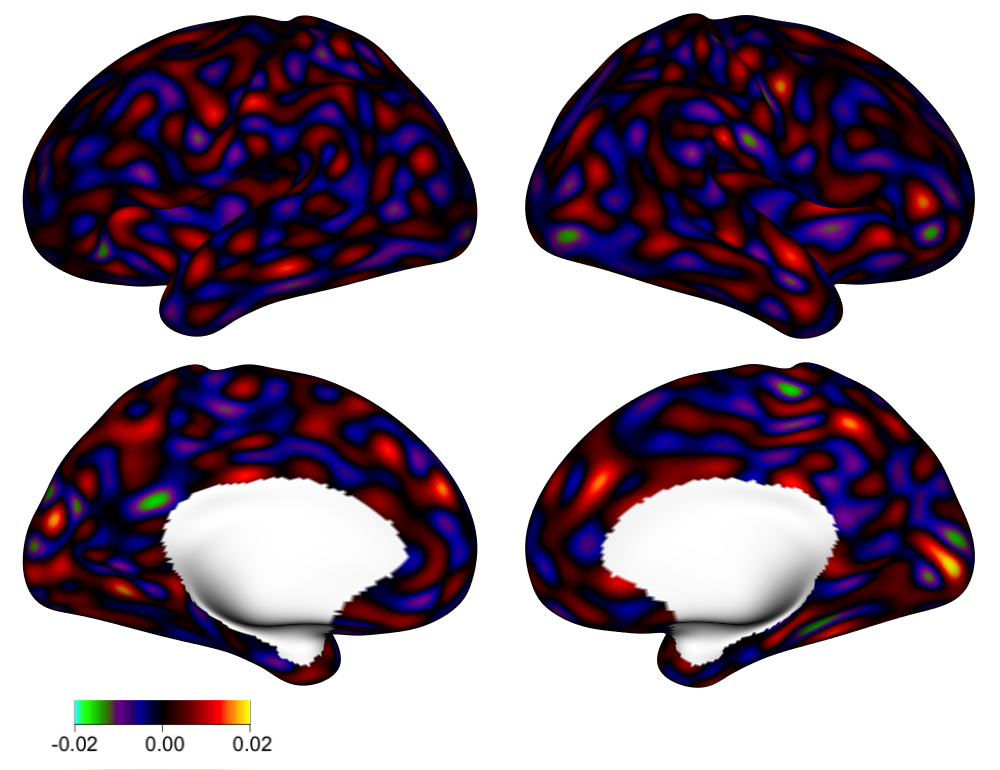}
\end{subfigure}
\begin{subfigure}{0.45\textwidth}
\caption{GPR + Horseshoe}
\includegraphics[width=1\linewidth]{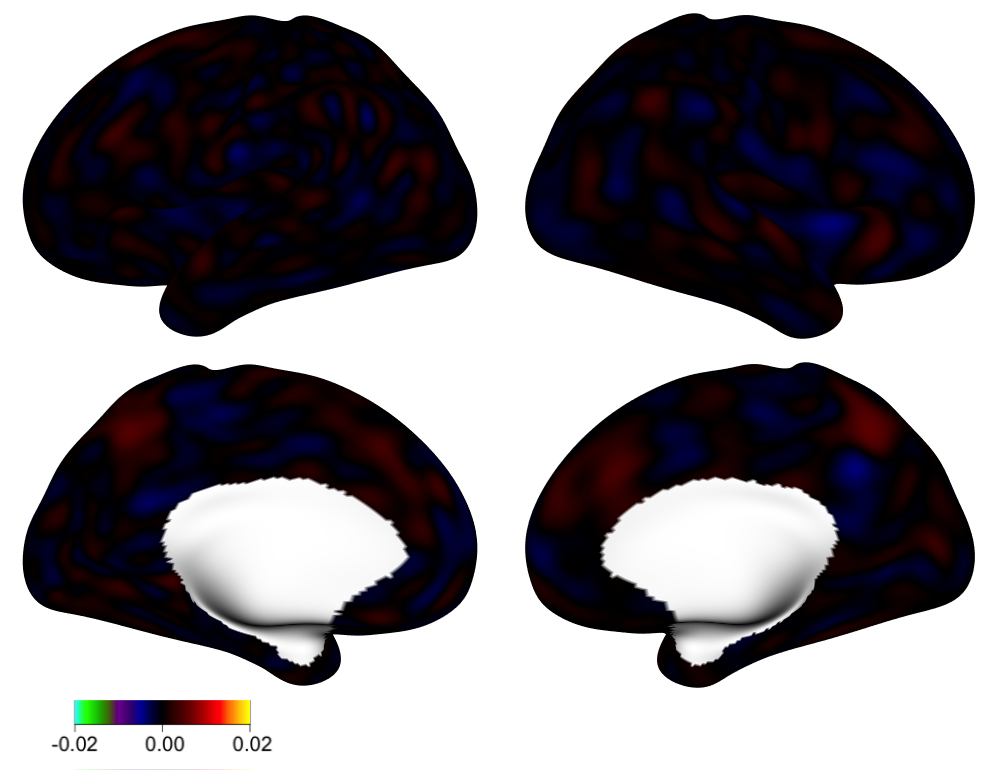}
\end{subfigure}

\begin{subfigure}{0.45\textwidth}
\caption{BR + Normal}
\includegraphics[width=1\linewidth]{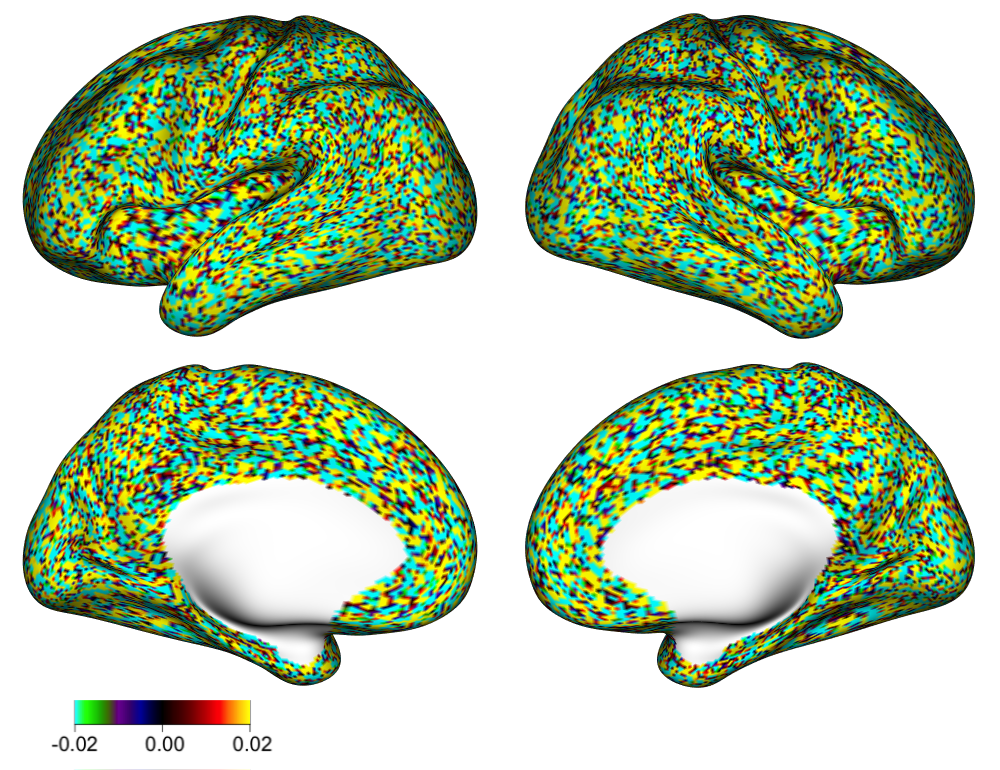}
\end{subfigure}
\begin{subfigure}{0.45\textwidth}
\caption{BR + Horseshoe}
\includegraphics[width=1\linewidth]{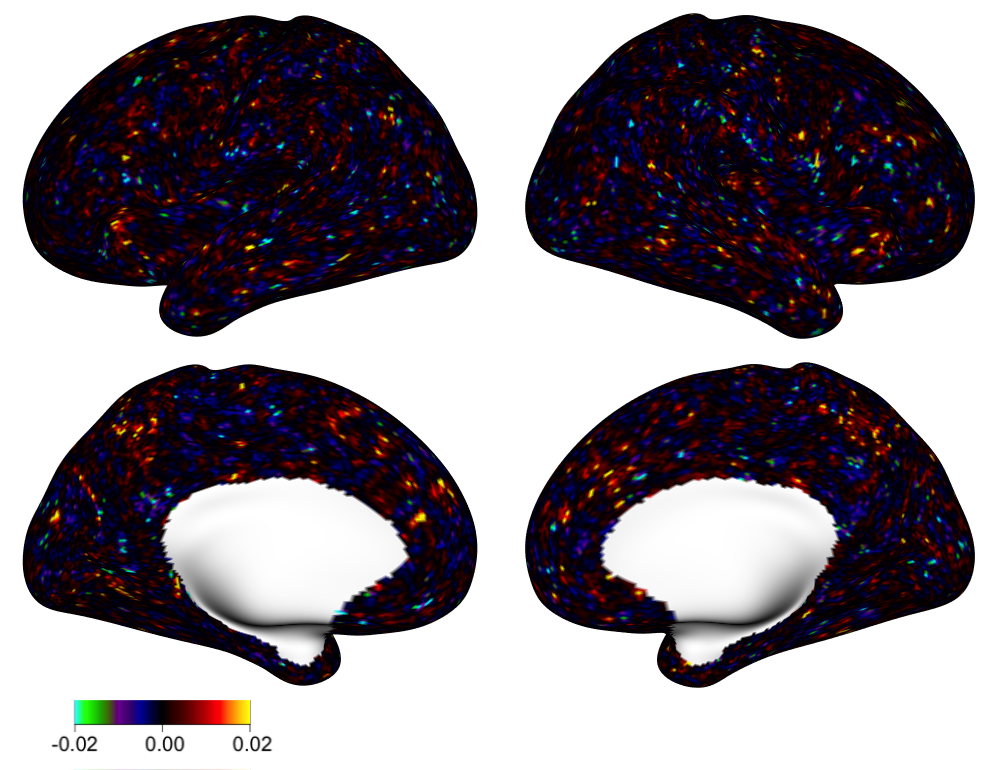}
\end{subfigure}

\begin{subfigure}{0.45\textwidth}
\caption{Ridge}
\includegraphics[width=1\linewidth]{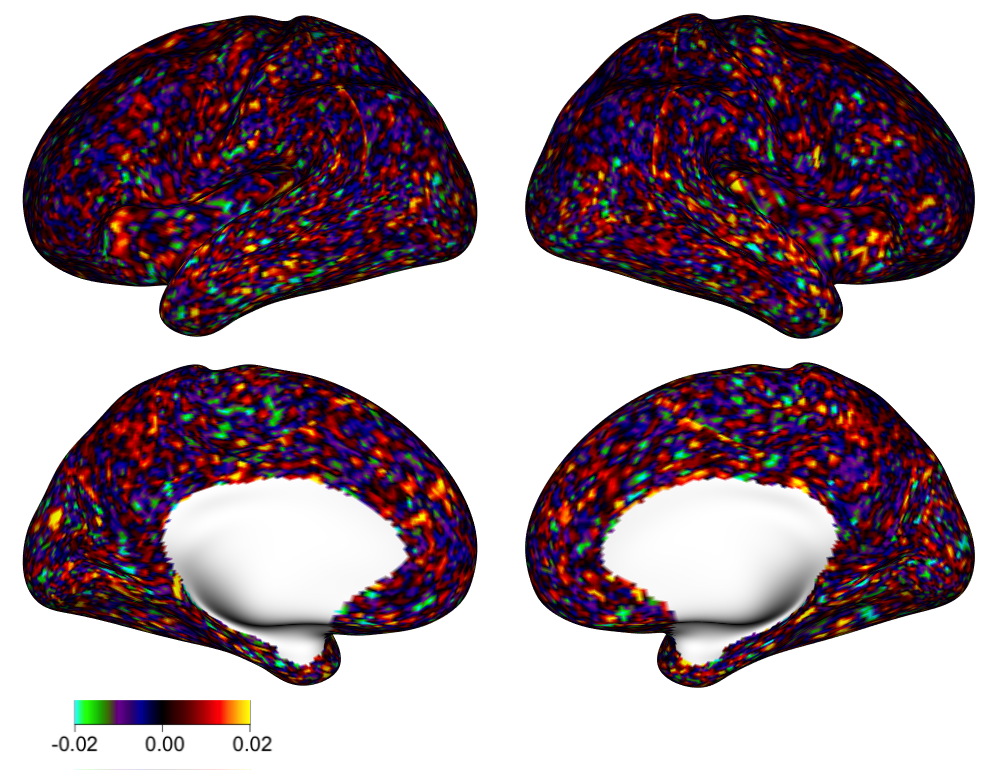}
\end{subfigure}
\begin{subfigure}{0.45\textwidth}
\caption{LASSO}
\includegraphics[width=1\linewidth]{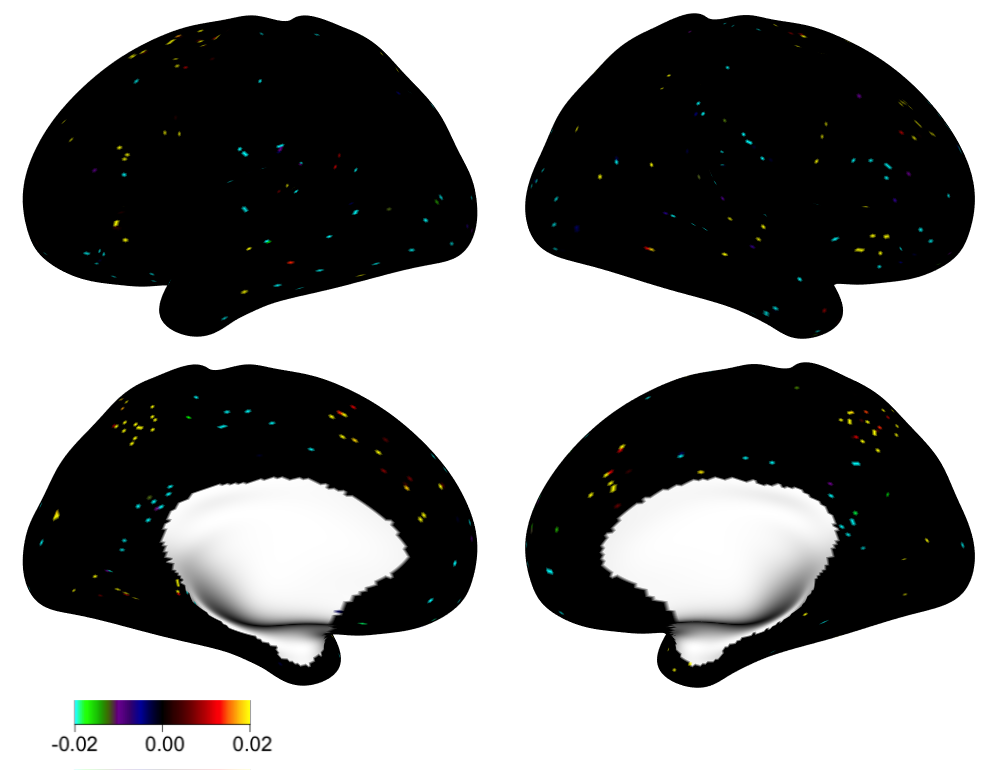}
\end{subfigure}
\end{center}
\end{adjustbox}

\caption[ABCD application: Comparison of the estimated parameter maps from our RTGP model and other baseline models.]{Comparison of the estimated parameter maps from (a)-(b) our RTGP model (unthresholded and thresholded) and (c)-(h) the other baseline approaches. Note that the we limit the colourbar to the range of the RTGP parameter values which lie between -0.02 to 0.02 for a fair comparison of models. However, the (f) BR + Horseshoe model and (h) the frequentist LASSO regression both exhibit far more extreme values which lie outside the plotted range.}
\label{fig: ch4-abcd-results-beta}
\end{figure}

The analysis of fMRI data in the ABCD study reveals similar trends as the simulation studies for the parameter estimation of the spatially varying regression coefficients in the scalar-on-image regression associating composite intelligence scores with the emotional n-back task fMRI test statistic values while accounting for confounding variables. Figure~\ref{fig: ch4-abcd-results-beta} shows that the Bayesian models that directly place a prior on the image coefficients, BR + Normal and BR + Horseshoe, or their frequentist counterparts, Ridge and LASSO regression, do not account for any spatial dependence in the model specification and therefore yield speckled results. Moreover, BR + Horseshoe and LASSO models do exhibit a certain level of shrinkage behaviour; however, neither models are able to recover any spatial clusters and exhibit speckled, extreme parameter estimation across the cortex. The GPR + Normal model exhibits more spatially smooth parameter estimates but does not seem to be sensitive enough to pick up any signal. For the other Gaussian process regression with a Horseshoe prior on the basis coefficients the effect size map shown in Figure~\ref{fig: ch4-abcd-results-beta} (d) depicts smooth parameter estimates across the cortex but exhibits a far smaller ranger of estimated parameter values than the other approaches with values between -0.004 and 0.004 which suggests that the global variance parameter of the Horseshoe prior biases the parameter estimates that should have a larger signal to a smaller effect size. On the other hand, our model RTGP showcases the benefit of thresholding the Gaussian process prior which is placed over the spatially varying image coefficients where the image coefficients drawn from a GP prior are thresholded to 0 if the latent variable placed at each vertex does not surpass a estimated threshold. The benefit being that parameter estimation and variable selection occurs separately. Hence, the parameter map in Figure~\ref{fig: ch4-abcd-results-beta} (b) of our RTGP model does not suffer from the same problem as the GPR + Horseshoe model in Figure~\ref{fig: ch4-abcd-results-beta} (d) where the global variance has the power to bias all coefficients in favour of stronger shrinkage. 

\begin{figure}[H]
\begin{adjustbox}{minipage=\linewidth,scale=1}
\begin{center}

\begin{subfigure}{0.45\textwidth}
\caption{RTGP}
\includegraphics[width=1\linewidth]{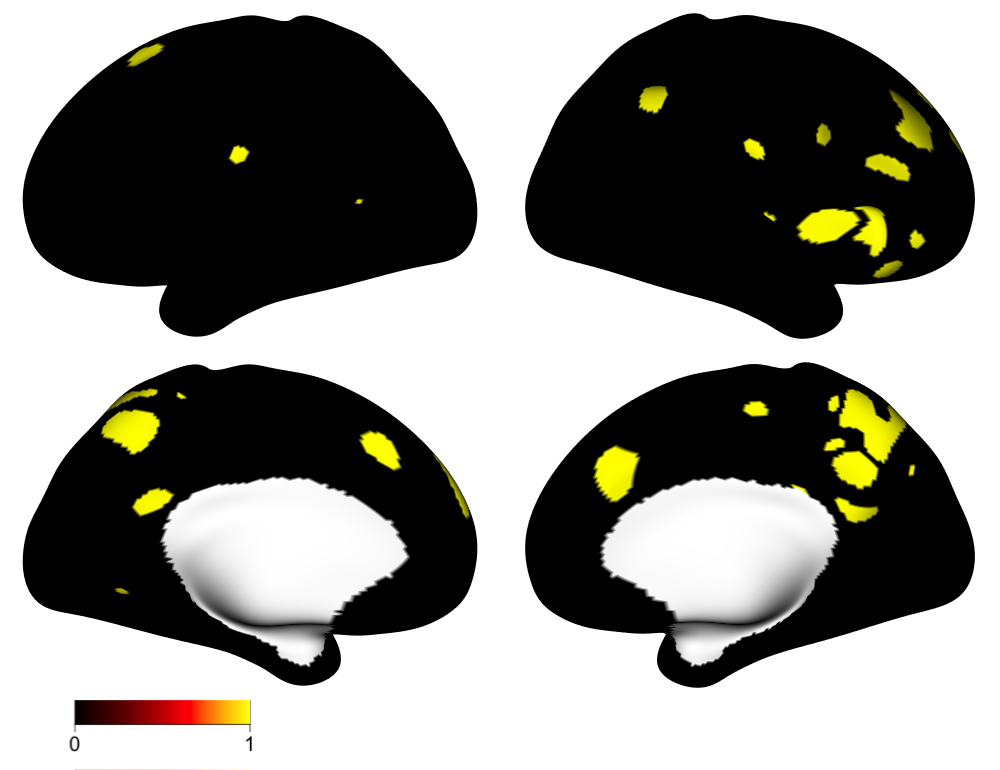}
\end{subfigure}
\begin{subfigure}{0.45\textwidth}
\caption{GPR + Normal}
\includegraphics[width=1\linewidth]{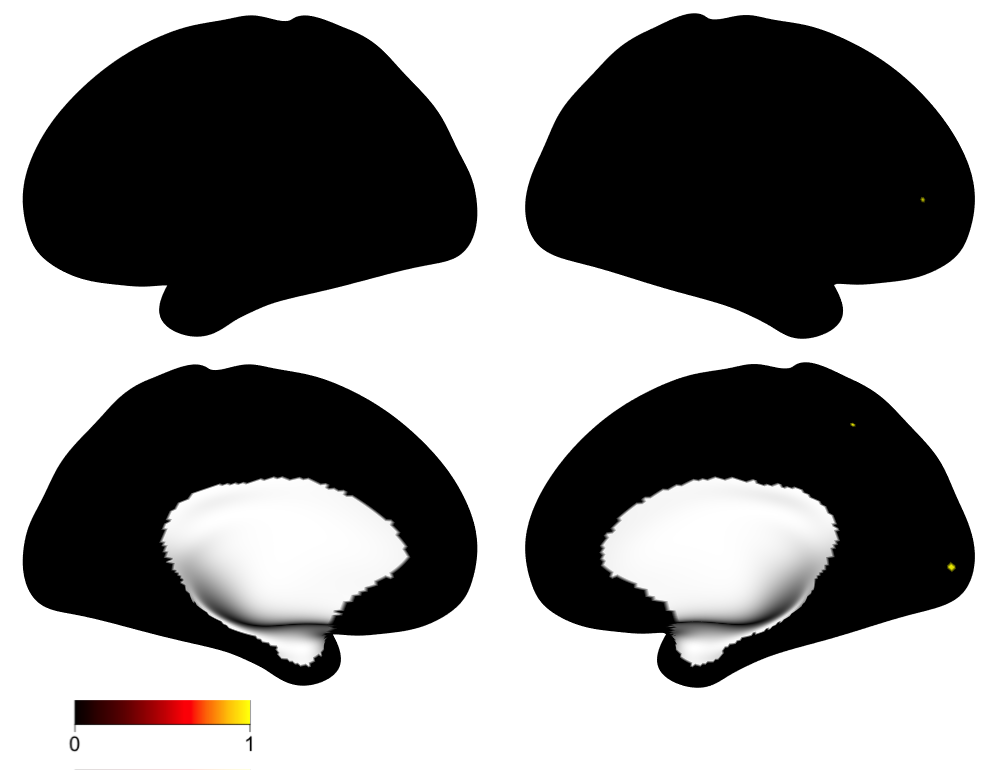}
\end{subfigure}

\begin{subfigure}{0.45\textwidth}
\caption{GPR + Horseshoe}
\includegraphics[width=1\linewidth]{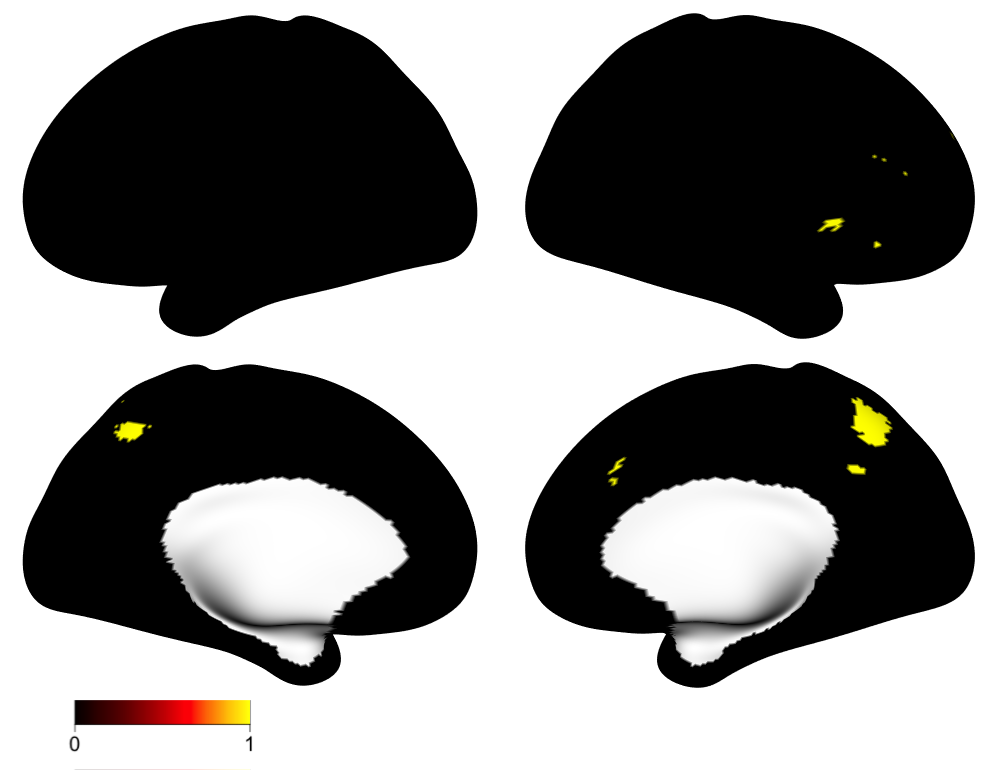}
\end{subfigure}
\begin{subfigure}{0.45\textwidth}
\caption{Overlap: RTGP \& GPR + Horseshoe}
\includegraphics[width=1\linewidth]{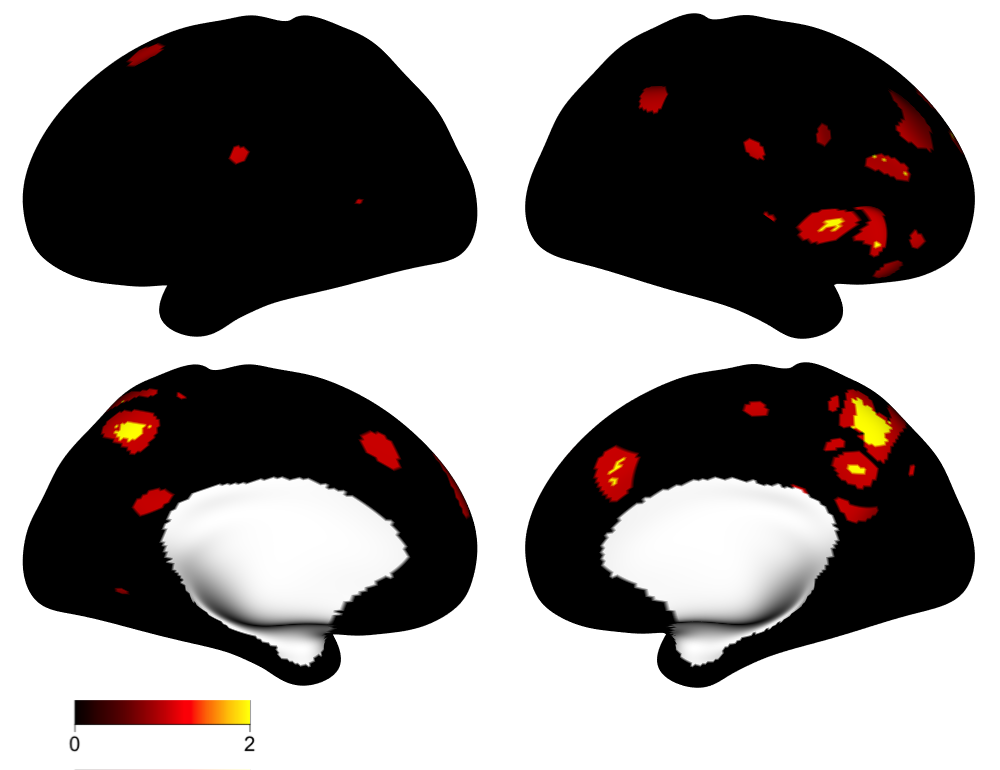}
\end{subfigure}
\end{center}
\end{adjustbox}

\caption[ABCD application: Comparison of the estimated binary significance maps from our RTGP model and other baseline models.]{Comparison of the estimated binary significance maps from (a) our RTGP model to the other baseline approaches, (b) GPR + Normal and (c) GPR + Horseshoe, and (d) overlapping the significance results of the RTGP model (red) and GPR + Horseshoe model (yellow). Note that the other models are omitted as significance can either not be determined (Ridge), significance is too speckled (LASSO) or no vertex is significant in the estimated map (BR + Normal, BR + Horseshoe).}
\label{fig: ch4-abcd-results-significance}
\end{figure}

For variable selection, we determine binary significance in the same manner as described in Section~\ref{sec: results} where we threshold the expected value $\mathds{E}[I(|\alpha(s_j)| > \delta)]$ for every vertex $j=1,\dots,M$ with 0.5 for the RTGP model and in the GPR + Normal and GPR + Horseshoe model we determine whether or not 0 is included in the 95\%-HPDI interval. We acknowledge that presently there is no principled way to find a significance threshold that corresponds to the same level of sensitivity versus specificity trade-off across models; however, we show that for the RTGP model varying the significance threshold yields comparable variable selection results and a similar level of out-of-sample predictive MSE across the thresholds, see Web Appendix A, Section~6.6. We also find that varying the size of the HPDI for the GPR baseline models does not drastically change the significance results. Figure~\ref{fig: ch4-abcd-results-significance} suggests that our RTGP model potentially produces more sensitive results than the GPR + Horseshoe baseline approach which yields significant results at the same vertices but has far less spatial activation than the RTGP approach, see Figure~\ref{fig: ch4-abcd-results-significance} (d) where we plot the overlap between the two binary significance maps and display the RTGP activation in red and any overlap between the approaches in yellow. Our RTGP model finds that 2,310 vertices are significant, the GPR + Normal model results in only 9 activations and the GPR + Horseshoe approach yields 247 significant vertices. 

Unfortunately, the problem of associating intelligence scores with the emotional n-back task fMRI has only been studied as an image-on-scalar problem which does not provide us with an indication of where we might expect true significance. We therefore can only suggest that our RTGP model has the potential to uncover significance of smaller effects yielding more sensitive results in data with complex spatial structures whereas the baseline approach GPR + Horseshoe provides potentially more conservative results in this real data application. Hence, the researcher can use either approach for prediction but should carefully evaluate whether or not they desire small areas of strong effect as shown by GPR + Horseshoe or potentially also consider subtler effects that can be relevant towards the prediction as additionally discovered by RTGP. In the case of RTGP, approximately the same areas of the brain in the dorsolateral and medial prefrontal cortices and precuneus bilaterally were found to have strong positive associations relevant towards the prediction of intelligence scores as the image-on-scalar regression studying the reverse problem \citep{Makowski2023}. In Web Appendix A, Section~6.7, we show that this result is robust to removing 10\% of the training data and performing 10-fold cross-validation on various splits of the training data. Lastly, Table~\ref{tab: ch4-results-abcd-prediction} also shows that the only two models that have reasonable variable selection results in addition to low predictive MSE and high R$^2$ are GPR + Horseshoe and our suggested approach RTGP. 

\begin{table}[]
    \centering
\begin{tabular}{l|rrrr}  \hline \textbf{Left Hemisphere} & R$^2$ (train) & MSE (train) & R$^2$ (test) & MSE (test) \\   
\hline 
Ridge & 66.66 & 110.9729 & 18.33 & 252.4871 \\ 
  LASSO & 35.45 & 188.6745 & 28.34 & 216.8615 \\   
  BR + Normal & \textbf{100.00} & \textbf{0.0201} & 4.42 & 532.4117 \\ 
  BR + Horseshoe & 49.95 & 149.9042 & 27.48 & 218.8589 \\   
  GPR + Normal & 26.25 & 222.9945 & 16.01 & 286.7683 \\ 
  GPR + Horseshoe & 31.09 & 198.3787 & \textbf{29.97} & \textbf{211.6691} \\   
  RTGP& 27.87 & 208.9946 & 29.96  & 211.8090\\
    \hline
    \textbf{Right Hemisphere} & R$^2$ (train) & MSE (train) & R$^2$ (test) & MSE (test) \\   
    \hline 
    Ridge & 66.74 & 109.7410 & 19.35 & 246.4008 \\ 
  LASSO & 36.18 & 186.8528 & 28.43 & 216.6652 \\   
  BR + Normal & \textbf{100.00} & \textbf{0.0067} & 2.63 & 526.8691 \\ 
  BR + Horseshoe & 52.00 & 144.3401 & 24.88 & 227.0925 \\   
  GPR + Normal & 26.74 & 222.8687 & 15.18 & 292.7009 \\ 
  GPR + Horseshoe & 31.59 & 196.8422 & 28.39 & 215.9624 \\     
  RTGP & 29.91 & 203.1902 & \textbf{29.27} & \textbf{215.5252} \\
    \hline
    \vspace{0.5cm}
    \end{tabular}
    \caption{Comparison of prediction results of ABCD study analysis evaluated with training and test R$^2$ (in \%) and predictive MSE for the left and the right hemisphere of the brain. The models which do not take any spatial dependence between the vertices into account exhibit overfitting and perform poorly in the test dataset. The GPR + Horseshoe and RTGP possess on par predictive performance.}
    \label{tab: ch4-results-abcd-prediction}
\end{table}

\section{Discussion and Future Work}
\label{sec: discussion}

Within this work we have proposed a Bayesian nonparameteric scalar-on-image regression model with a relaxed-thresholded Gaussian process prior. Our contributions are three-fold: 1) we introduce a more flexible thresholding prior that is able to perform hard- and soft-thresholding of spatially varying coefficients alike, 2) we increase the scalability of our scalar-on-image regression models with a thresholded GP prior to large-scale applications by using a variational approximation to the posterior instead of MCMC sampling, and 3) to the best of our knowledge the Karhunen-L\`oeve expansion has not been used on kernels with a correlation function that measures distance on the surface. 

We do note that our model relies on an approximation of GPs, specifically the Karhunen-L\`oeve expansion, which is reliant on determining a number of basis functions that is far lower than the number of vertices of the image. The goal of the approximation is to yield an adequate approximation while being computationally much faster. We determine this number by satisfying a certain percentage of total variation, see Equation~\ref{eq: percentage_total_variation} and Web Appendix A, Section~6.3, and find that capturing only 30\% of the total variation yields adequate results. We do not claim that this holds for other applications and hence our VI approach would suffer from the same computational bottleneck as in MCMC sampling if the number of basis functions is increased to capture a value closer to 100\% of the total variation.

Regarding future work, the current scalar-on-image regression method can be extended to multiple data modalities, such as structural, diffusion-weighted or resting-state functional MRI. Current approaches often summarise the results of different image modalities \citep{Kang2018}; however, introducing smoothness and sparsity for each modality separately would require a separate thresholded GP prior with its own thresholding parameter. The advantage of introducing a separate prior for each spatially varying coefficient set is that the image dimension would not be required to match and hence, enabling the combination of various image modalities as input sources. 

As another area of future work, we could explore other potential kernels for the covariance function of the GP prior on the spatially varying coefficients. Our simulation studies and real data application revealed that the GPR + Horseshoe model has a lower predictive MSE than RTGP in both scenarios. An interesting area of future work would hence be to test if a Horseshoe kernel can be used in lieu of the two parameter exponential radial basis function that we currently employ and achieve better performance with respect to prediction in addition to variable selection while still retaining a scalable inference algorithm.


\backmatter


\section*{Acknowledgements}

Thank you to Chris Holmes for his feedback on this work. AM is funded by \mbox{EPSRC} StatML CDT (EP/S023151/1), \mbox{Oxford}-Radcliffe scholarship and Novartis, TEN by the Wellcome Trust (100309/Z/12/Z) and NIH grant 1R01DA048993, TDJ by NIH R01DA048993, and JK by NIH R01DA048993 and NIH R01MH105561. We would also like to thank Mike Angstadt and Chandra Sripada for sending a preprocessed collection of CIFTI files from the ABCD study. \vspace*{-8pt}

\section*{Data Availability Statement}
The ABCD Study data is publicly available and can be downloaded from the NIMH Data Archive (\url{https://nda.nih.gov/}) for approved users.


%
%
\bibliographystyle{biom} 
\bibliography{biomsample_bib}


\section*{Supporting Information}

Web Appendix A, referenced throughout this paper, is available at \url{https://drive.google.com/file/d/1SNS0T6ptIGLfs67zYrZ9Bz0-DgzCIRgz/view?usp=sharing}. We also provide all code on our Github (\url{https://github.com/annamenacher/RTGP}).\vspace*{-8pt}

\appendix




\label{lastpage}

\end{document}